\documentstyle[aaspp4,11pt,psfig]{article}

\begin{document}

\def\spose#1{\hbox to 0pt{#1\hss}}
\def\simlt{\mathrel{\spose{\lower 3pt\hbox{$\mathchar"218$}}
     \raise 2.0pt\hbox{$\mathchar"13C$}}}
\def\simgt{\mathrel{\spose{\lower 3pt\hbox{$\mathchar"218$}}
     \raise 2.0pt\hbox{$\mathchar"13E$}}}

{THE LYMAN ALPHA FOREST IN THE SPECTRA OF QSOS}\\
\bigskip

\noindent Michael Rauch\\
Astronomy Department, California Institute of Technology\\
Pasadena, CA 91125, USA\\
mr@astro.caltech.edu
\medskip

\medskip


\medskip

\pagebreak

\centerline{\bf Abstract}
Observations of redshifted Lyman $\alpha$ forest absorption in the
spectra of quasistellar objects (QSOs) provide a highly sensitive probe of the distribution of
gaseous matter in the universe.  Over the past two decades optical
spectroscopy with large ground-based telescopes, and more
recently ultraviolet spectroscopy from space have yielded a wealth of
information on what appears to be a gaseous, photoionized
intergalactic medium, partly enriched by the products of stellar
nucleosynthesis, residing in coherent structures over many hundreds
of kiloparsecs.

Recent progress with cosmological hydro-simulations based on
hierarchical structure formation models has led to important insights
into the physical structures giving rise to the forest.
If these ideas are correct, a truely inter- and proto-galactic medium
[at high redshift ($z \sim 3$), the main repository of baryons] collapses
under the influence of dark matter gravity into flattened or
filamentary structures, which are seen in absorption against 
background QSOs. With decreasing redshift, galaxies forming in the denser
regions, may contribute an increasing part of the Ly$\alpha$ absorption
cross-section. Comparisons between large data samples from the 
new generation of telescopes and artificial Ly$\alpha$ forest spectra from
cosmological simulations promise to become a useful cosmological tool.

\pagebreak

\section{\large INTRODUCTION}

The Lyman $\alpha$ forest is an absorption phenomenon in the spectra of
background quasistellar objects (QSOs). It can be observed in the ultraviolet
(UV) and optical wavelength
range, from the local universe up to the highest redshifts where QSOs
are found (currently z$\sim$ 5).  Neutral hydrogen intersected by the
line of sight (LOS) to a QSO will cause absorption of the QSO continuum
by the redshifted Ly$\alpha$ (1215.67 \AA ) UV resonance line.  In
an expanding universe homogeneously filled with gas, the continuously
redshifted Ly$\alpha$ line will produce an absorption trough blueward
of the QSO's Ly$\alpha$ emission line (independent predictions by Gunn
\& Peterson 1965; Scheuer 1965; Shklovski 1965).  Gunn \& Peterson
found such a spectral region of reduced flux, and used this measurement
to put upper limits on the amount of intergalactic neutral hydrogen.
The large cross-section for the Ly$\alpha$ transition makes this
technique by far the most sensitive method for detecting baryons at any
redshift.

Bahcall \& Salpeter (1965) suggested  that there should also be a 
population of discrete absorption lines from a more clumpy gas
distribution, specifically from intervening groups of galaxies.
Discrete lines were observed shortly thereafter (Lynds \&
Stockton 1966; Burbidge et al 1966; Stockton \& Lynds
1966; and Kinman 1966), but the quest for their precise origin has
given rise to a long and, at times, controversial debate; only in
recent years does the issue appear to have been resolved (see below). Soon
thereafter,
the simultaneous detection of higher order lines of the Lyman
series (e.g., Baldwin et al. 1974) had confirmed the suggestion (Lynds
1970) that most of the absorption is indeed from HI Ly$\alpha$. At higher spectral resolution, the Ly$\alpha$ forest can be
resolved into hundreds (in z $> 2$ QSO spectra) of distinct absorption
lines, the appearance of which gave rise to the label Ly$\alpha$
forest (Weymann et al 1981); see Figure 1.  A small fraction of the
lines hidden in the forest are not caused by HI but belong to UV
transitions from several common metal or heavy element ions (various
ionization stages of C, O, Mg, Si, Fe and Al are most frequently seen).
These metal lines are invariably associated with strong Ly$\alpha$
lines. At column densities N(HI) exceeding 10$^{17}$ cm$^{-2}$ the gas
becomes optically thick to ionizing radiation and a discontinuity at
the Lyman limit (912 \AA ) is detectable. In systems with N(HI) larger
than $\sim 10^{19}$ cm$^{-2}$, selfshielding renders the gas
predominantly neutral. The damping wings of the Lorentzian component of
the absorption profile beginn to be detected from about the same column
density, reaching their maximum in the ''damped Ly$\alpha$ systems''.

The question of whether
the majority of the absorption systems are truly intervening at
cosmological distances from the quasar, or ejected by it, which had
received considerable interest in the earlier days, is now settled in
favor of the intervening hypothesis.  The huge momentum requirements
for ejection (Goldreich \& Sargent 1976), the outcome of the Bahcall \&
Peebles test (1969; Young et al 1982a) for a random
redshift distribution of absorbers to different QSOs, the discovery of
galaxies at the same redshifts as metal absorption systems (Bergeron
1986), and the detection of high metallicity gas in systems close to
the QSO and low metallicities more than 30000 km$^{-1}$ away from it
(Petitjean et al 1994) leave no doubt that most of the
systems are not physically related to the QSO against which they are
observed.

The basic observational properties of the Ly$\alpha$ forest were
established in the late 1970s and early 1980s when the combination of
4m telescopes (e.g., the AAT, KPNO, MMT, Palomar) and sensitive photon
counting electronic detectors (e.g. the University College London's
IPCS) first permitted quantitative spectroscopy on high redshift QSOs
to be performed.  Making use of the new technology the work by Sargent
et al. (1980) set the stage for what for many years has been the
standard picture of the Ly$\alpha$ forest: Ly$\alpha$ absorbers were
found to be consistent with a new class of astronomical objects,
intergalactic gas clouds, which are distinct from galaxies (and metal absorption
systems) by their large rate of incidence (d{\cal N}/dz) and their weak
clustering. Upper limits on the gas temperature and estimates for the
ambient UV flux and for the cloud sizes were found to be consistent
with a highly ionized ($n_{HI}/n_{H} \leq 10^{-4}$) optically thin gas
kept at a temperature $T \sim 3\times 10^4 K$ by photoionization
heating.  Sargent et al (1980) suggested that denser clouds in pressure
equilibrium with a hotter (ie. more tenuous) confining intercloud
medium (ICM) could explain the apparent lack of change of these objects with
time. It was argued that this picture matches the inferred cloud
properties better than clouds held together by gravity, and there were
a number of other appealing features.  In the wake of the dark matter-based structure formation scenarios, the pressure confined clouds have
given way to models where Ly$\alpha$ clouds arise as a natural
immediate consequence of gravitational collapse. These results are
discussed later.

In an earlier review, Weymann et al (1981) introduced a
classification of absorption systems that is still useful, although
some of the distinctions introduced have been blurred by the most recent research
(Tytler et al 1995; Cowie et al 1995). In particular, the earlier
review distinguished two classes of absorption systems, physically
separated from the QSO environment, according to whether they do, or do
not, show metal absorption lines in addition to the ubiquituous
Ly$\alpha$.  For most of the Ly$\alpha$ clouds detectable with current
technology (N(HI) $\simgt$ 10$^{12}$ cm$^{-2}$) metal lines with
metallicities common at high redshifts  ($Z\simlt$$10^{-2} $ $Z\odot$)
are simply below the detection threshold. Therefore this classification is simply
an observational one.  Rather than explore the nature of the division,
if appropriate, between metal absorbers and Ly$\alpha$ systems here we
will concentrate on the low column density absorbers.  The study of
metal absorption systems, possibly of great relevance to 
galaxy formation, is left for future review.

Below we first discuss observational techniques and observed
properties of Ly$\alpha$ systems (using the terms absorption
systems, absorbers, clouds interchangeably). Then we 
turn to various models of Ly$\alpha$ absorbers, and finally we address
some recent results, and speculate about future developments.

\section{\large OBSERVATIONAL APPROACHES}

\medskip

\subsection{\it Technical Possibilities, Observational Constraints}

Observational progress with traditional Ly$\alpha$ studies can
conveniently be charted in terms of two limiting factors: The spectral
resolution, and the signal-to-noise ratio.  The early (photographic)
spectroscopy in the 1960s typically had to rely on resolutions of 10
-- 20 \AA . At optical wavelengths this is barely sufficient to resolve
velocity dispersions characteristic of galaxy clusters. Later, in
particular the combination of echelle spectrographs and CCD detectors
permitted QSO spectroscopy to be done with 4m telescopes at resolutions as high as  R$\sim$ 5$\times$10$^4$.  QSO observers have been quick to make
use of these advances as is obvious from the semantic drift of 
``high resolution'': in 1979 it meant 0.8 \AA\ (Sargent et al.  1979),
in 1984, 0.25 \AA\ (Carswell et al. 1984), and in 1990, 0.08 \AA\ 
(Pettini et al.  1990). Naturally, the  average signal-to-noise ratio
per resolution element did not benefit from the increase in spectral
dispersion, and a single high resolution QSO spectrum required an
embarassingly large number of nights on a 4m telescope.  Thus various
ways of extracting information from the Ly$\alpha$ forest have been
developed in parallel.  Some were tailored to detailed analysis of
individual, expensive, high resolution spectra (line profile fitting),
while others could be applied more automatically to larger samples of
low resolution data (mean absorption; equivalent width measurements).
The 10m Keck telescope with its powerful High Resolution Spectrograph
(HIRES; Vogt et al.  1994) brought signal-to-noise ratios of $\sim$ 100 within reach of high
resolution (FWHM $\sim 7$ kms$^{-1}$) absorption line studies,
rendering some of the low resolution approaches obsolete. The new
limiting factor for large telescope spectroscopy is not resolution nor
collecting area but manpower -- coping (intelligently) with the
continuous stream of large datasets already, or soon, available from
the Keck telescopes, the Magellan, the Hobby-Eberly, the ESO-VLT, the
MMT, etc.  Recently, progress has also come from extending the
wavelength regime into the UV band with the Hubble Space Telescope (HST) and
the Hopkins Ultraviolet Telescope (HUT).  In particular, the advent of
the HST with its high resolution UV spectrographs
has helped to compensate to some extent for the fact that
optical Ly$\alpha$ spectroscopy can only sample the universe at redshifts
larger than $\sim 2.5$. We can now study the absorber
properties and the absorber-galaxy connection in the local universe,
and, at high redshift, the  far-UV (rest frame) helium Ly$\alpha$
forest.

\subsection{\it Low Resolution Spectroscopy: Mean Absorption}

The most basic observable of the Ly$\alpha$ forest
is the quantity originally sought by Gunn and Peterson (1965),
the flux decrement $D$, or the mean fraction of the QSO continuum
absorbed. 
It has become standard practice to
measure the mean flux decrement $D_A$ between the Ly$\alpha$ and Ly$\beta$
emission lines (Oke \& Korycansky 1982):

\begin{eqnarray}
D_A = \left\langle{1 - \frac{f_{obs}}{f_{cont}}}\right\rangle = \left\langle{1 - e^{-\tau}}\right\rangle = 1 - e^{-\tau_{\mathrm eff}}, \label{da}
\end{eqnarray}
where $f_{obs}$  is the observed (=residual) flux, $f_{cont}$ the
estimated flux of the unabsorbed continuum, and $\tau$ is the resonance
line optical depth as a function of wavelength or redshift.  The
absorption is measured against a continuum level usually taken to be a
power law in wavelength extrapolated from the region redward of the
Ly$\alpha$ emission line.  A knowledge of the mean absorption allows us, for example,
to determine the resulting broadband color
gradients in a QSO or high redshift galaxy spectrum, as caused by
the Ly$\alpha$ forest.  Measurements of $D_A$ from large datasets were performed by (among others) Oke \&
Korycansky (1982), Bechtold et al. (1984), O`Brian et al (1988), Steidel \& Sargent (1987), Giallongo \& Cristiani
(1990), Dobrzycki \& Bechtold (1991), Schneider et al
(1991, and refs. therein).  Obviously, at the price of getting only one
number out of each QSO spectrum, this technique gives the most
model-independent measurement possible.

With $D_A$ measurements available over a range of redshifts the redshift
evolution of the Ly$\alpha$ forest can be investigated.
Here the concept of an effective optical depth $\tau_{\mathrm eff}(z)$, as defined in
Equation (\ref{da}) becomes useful.
If we characterize a Ly$\alpha$ forest as a random distribution of
absorption systems in column density $N$, Doppler parameter $b$, and
redshift $z$ space, such that the number of lines  per interval $dN$, $db$ and $dz$ is given by ${\cal F}(N,b,z)dNdbdz$, then
\begin{eqnarray}
\tau_{\mathrm eff} = \int_{z_1}^{z_2}\int_{b_1}^{b_2}\int_{N_1}^{N_2}\left(1-e^{-\tau(N,b)}\right) {\cal F}(N,b,z) dN db dz
\end{eqnarray}
(Paresce et al 1980), for a population of absorbers without spatial or velocity correlations.
Assuming that the $N$ and $b$ distribution functions are independent of redshift,
and the redshift evolution of the number density of lines can be approximated by a power law, we can write ${\cal
F}(N,b,z) = (1+z)^{\gamma}F(N,b)$, and
\begin{eqnarray}
\tau_{\mathrm eff}(z) = \frac{(1+z)^{\gamma+1}}{\lambda_0}\int_{b_1}^{b_2}\int_{N_1}^{N_2}{F}(N,b) W(N,b) dN db \label{zuo}
\end{eqnarray}
where the rest frame equivalent width is given by $W =
(1+z)^{-1}\lambda_0^{-1}\int d\lambda \left(1-e^{-\tau}\right)$.

This relation enables us to measure the redshift evolution of the
number density of Ly$\alpha$ forest clouds, $d{\cal N}/dz \propto
(1+z)^{\gamma}$, from the redshift dependence of the effective
optical depth ($\tau_{\mathrm eff} \propto (1+z)^{\gamma+1}$) 
even if we do not resolve the individual absorption lines
(Young et al 1982b; Jenkins \& Ostriker 1991; Zuo 1993; 
Press et al 1993; Lu \& Zuo 1994).
The results of this approach are discussed below in the section
on Ly$\alpha$ forest evolution, together with the conclusions from
line counting methods.

The largest
uncertainties in $D$ or $\tau_{\mathrm eff}$ are probably caused by our ignorance
about the precise QSO continuum level, against which the absorption is
measured. Additional errors  stem from the amount of
absorption contributed by metal lines, which often cannot
be identified as such and then removed from low resolution Ly$\alpha$ spectra.

\bigskip

\subsection{\it Intermediate Resolution Spectroscopy: Line Counting}


At higher
resolution, where it becomes possible to distinguish between discrete
absorption lines, the distribution of the lines in terms of equivalent
width $W$ and redshift $z$ is the next most sophisticated tool. Just as the
mean absorption $D$ under certain conditions provides a measure of the
mean (gas) density of the universe, so does the number of lines per
unit equivalent width, $d^2{\cal N}/dWdz$ essentially
measure the clumpiness of the Ly$\alpha$ forest gas. An
exponential distribution in $W$ and a power law dependence on $(1+z)$ have been found to provide a reasonable match
to the observed line counts (Sargent et al (1980); Young et al (1982b); Murdoch et al (1986)). For lines above a rest
equivalent width threshold $W>$ 0.16 \AA ,
 \begin{eqnarray}
\frac{d^2\cal N}{dWdz} = \frac{W}{W_*} e^{-W/W_*}(1+z)^{\gamma}, \label{ew}
\end{eqnarray}
with a typical $W_* \approx$ 0.27 \AA\ (Bechtold 1994), and  $1.5<\gamma<3$
(see also the discussion on redshift evolution below).
The multiplicative form of Equation (\ref{ew}) is justified by a relatively  weak
dependence of $W_*$ on $z$ (Murdoch et al 1986).
The exponential model fits less well for lines with $W<0.3$ \AA\ as
the weaker lines are moving off the saturated part of the curve of
growth and become more numerous.

Unfortunately, statistics involving the observed equivalent width distribution
are not easy to interpret in physical terms. The $W$ values are usually
obtained by simply measuring statistically significant downward
excursions from the QSO continuum. Without properly deblending the
lines (impossible at low resolution) the curve of growth cannot be used
to relate $W$ to the more meaningful parameters $N$ and $b$.

\subsection{\it High Resolution Spectroscopy: Voigt Profile Decomposition}

If the  Ly$\alpha$ forest is seen as an assemblage of redshifted lines
the standard arsenal of notions and techniques from stellar
spectroscopy becomes applicable.  For lower resolution data, the
equivalent width provides a combined measure of line width and
strength. In high resolution spectra (FWHM $< 25$ kms$^{-1}$) where the
typical Ly$\alpha$ line is resolved, the line shapes are found to be
reasonably well approximated by Voigt profiles (Carswell et al 1984).  Then line
width (Doppler parameter $b$ ($=\sqrt{2}\sigma)$), column density
$N(HI)$, and redshift $z$ of  an absorption line are the basic observables. The
statistics of the Ly$\alpha$ forest from high resolution studies largely have
been cast in terms of the distribution functions of these three
quantities and their correlations. The main advantage of the high
resolution approach is the opportunity of determining the shape of these
distribution functions without parametric prejudices, by directly
counting lines with parameters in a certain range.

The standard approach to Voigt profile fitting (Webb 1987; Carswell et
al. 1987) relies on $\chi^2$ minimization to achieve a complete
decomposition of the spectrum into as many independent Voigt profile
components as necessary to make the $\chi^2$ probability consistent
with random fluctuations. 
For stronger Ly$\alpha$ lines the higher order Lyman lines can provide
additional constraints when fitted simultaneously.
The absorption lines are measured against a
QSO continuum estimated locally from polynomial fits to spectral
regions deemed free of absorption.  A local high order continuum fit
(as compared to a global extrapolation with a physical model for the
QSO continuum) is necessary because the spectra are patched together
from many individual echelle orders with strong variations in
sensitivity. These variations do not divide out completely when dividing by the
flux of a standard star because the light going through a slit narrow
enough to ensure  slit-width limited resolution varies with the
seeing conditions and with the position of the object on the slit.
When applying a local fit to the continuum the
zeroth order contribution tends to be underestimated, i.e., the continuum
is drawn too low, which is the main drawback of this method.

Given sufficient spectral
resolution, and assuming that Ly$\alpha$ clouds are discrete
entities (in the sense of some of the models to be discussed below) the
profile fitting approach is the most physically meaningful way of
extracting information from the Ly$\alpha$ forest.  If the absorber is
a gas cloud with a purely Gaussian velocity dispersion (a thermal
Maxwell-Boltzmann distribution, plus any Gaussian contributions from
turbulence) a Voigt profile provides an exact description of the
absorption line shape. The Doppler parameter can then be written as the
quadratic sum of its individual contributions:
\begin{eqnarray}
b=\sqrt{\frac{2kT}{m} + b_{turb}^2}
\end{eqnarray}
Unfortunately, in most more realistic models of the absorbing gas
finite velocity and density gradients invalidate the assumptions 
underlying Voigt profile fitting, and the line parameters may have
less immediate physical meaning. 
Departures of the absorption line shape from a Voigt profile may
contain valuable information about the underlying nature of the
absorption systems, and different scenarios may have quite different
observational signatures.  Rotational motion (Weisheit 1978; Prochaska
\& Wolfe 1997), gravitational collapse ( McGill 1990; Meiksin 1994;
Rauch 1996) and galactic outflows (Fransson \& Epstein 1982; Wang 1995)
have been discussed in terms of the likely absorption line shapes they produce.
As yet, the quantitative application of these results has proven difficult,
because of the lack of realistic prototypical models for the actual
line formation, the rather subtle departures from Voigt profiles expected, and the
wide variety of profiles actually encountered. 

Non-Voigt profiles can still be fitted as blends of several Voigt
profiles, but the information about the non-thermal motion is encoded
in the spatial correlations between the individual profiles (Rauch
1996). Also, there is no guarantee that the number of components
necessary for a good fit converges with increasing signal-to-noise ratio. 
Clearly, for more general line formation models, global techniques of extraction the velocity information may be more
appropriate.

\section{\large OBSERVATIONAL RESULTS}

\subsection{\it Redshift Evolution of the Lyman Alpha Forest}

In an individual QSO line of sight, observations of the high redshift
($z\sim 3$) Ly$\alpha$ forest  can extend  over a redshift range
$\Delta z$ greater than unity. Then we are sampling a significant
fraction of a Hubble time, and it is natural to expect to see changes
in the absorption pattern, e.g., in the rate of incidence of absorption
lines with redshift, or in the mean optical depth.

\medskip

{\small
\noindent \it EVOLUTION OF THE LINE DENSITY \ \ } 
An analytic expression
(Wagoner 1967, Bahcall \& Peebles 1969) can be given for the number of
absorption systems per unit redshift, $d{\cal N}/dz$, in terms of the
comoving number density $n_0(z)$ of absorbers, the geometric absorption
cross section $\sigma(z)$, and the Hubble constant $H_0$:
\begin{eqnarray}
\frac{d{\cal N}}{dz} = \frac{c n_0(z) \sigma(z)}{H_0} \frac{1+z}{(1+q_oz)^{1/2}}.
\end{eqnarray}
For absorbers with no intrinsic evolution, 
\begin{eqnarray}
\frac{d{\cal N}}{dz} \propto \left\{\begin{array}{ll}
1+z,& q_0 = 0 \\
(1+z)^{1/2},& q_0 = \frac{1}{2}\end{array}\right..
\end{eqnarray}
The transformation from redshift $z$ to the coordinate 
$X(z) = \int_0^z (1+z)^{-1}(1+2q_0z)^{1/2} dz  \label{capx}$ (Wagoner 1967) is sometimes
used to
take out the mere cosmological redshift dependence,
such that $d{\cal N}/dX$ = $c n_0(z) \sigma(z) H_0^{-1}$.

\medskip

Peterson (1978) first pointed out an increase in the number of
Ly$\alpha$ clouds with redshift beyond what was expected for a
populations of non-evolving objects.  At first this result was subject
to some debate (see the summary by Murdoch et al. 1986), but it is now
clear that the Ly$\alpha$ forest as a whole evolves quite strongly with
$z$.  The observationally determined evolution in the number of
absorbers above a certain column density threshold is usually expressed
in the form
\begin{eqnarray}
\frac{d{\cal N}}{dz} = \left(\frac{d{\cal N}}{dz}\right)_0 (1+z)^{\gamma} 
\end{eqnarray}
(Sargent et al 1980; Young et al 1982b), where the
exponent includes the cosmological dependence.  Much observational
effort has been devoted to studying the redshift number density
evolution, but unfortunately the resulting conclusions are far from
robust. This is because Ly$\alpha$ cloud column densities $N$
are  distributed according to a power law $N^{-\beta}$ with index
$\beta \sim 1.5$ (see below) so the majority of lines
in any column density limited sample are always close to the threshold,
and small variations in imposing the threshold can cause large changes
in the estimated numbers of lines, and in $\gamma$.  Moreover, line
blending, and its dependence on spectral resolution, data quality, and
redshift can make a large difference in the normalization $(d{\cal
N}/dz)_0$, with individual studies differing by a factor of two or more
(the interested reader may refer to Parnell \& Carswell 1988; Liu \& Jones 1988; Trevese et al 1992, and Kim et al 1997 for a discussion of
blending effects).  Each individual study has a finite redshift range
available so that the uncertainty in the normalization is correlated with
the exponent $\gamma$; the differences between the observed values
of $\gamma$ exhibit a disturbingly large scatter. Thus the line counting
approach is unsatisfactory when it comes to redshift evolution. 
Rather than discussing the many individual (and sometimes
mutually inconsistent) contributions made to this question we will outline
some typical results as follows. 

At lower resolution (FWHM $\sim$ 50-100 kms$^{-1}$) large samples of lines
have been used to investigate this topic. The $\gamma$ values tend to
lie between the low value  $\gamma = 1.89\pm0.28$ obtained by Bechtold
(1994), ($W>$0.32 \AA ), and the high one $\gamma = 2.75\pm0.29$ (for
$W>$0.36 \AA ) from the study by Lu et al (1991).  High
resolution spectra, usually confined to $z>2$, tend to give an equally
wide range of values:  $\gamma = 1.7\pm 1.0$ ($W>$0.2 \AA ; Atwood et al 1985); $2.9\pm 0.3$ ($2<z<4.5$; Cooke et al. 1997);
$\gamma = 2.78\pm 0.71$ ($\log N > 13.77$; $2<z<3.5$; Kim et al 1997).
At $z>4$ the evolution appears to be accelerating, with $\gamma$ increasing
from a value just below 3 to 5.5 (Williger et al 1994).

\medskip

{\small
\noindent \it EVOLUTION OF THE MEAN ABSORPTION \ \ } 
Similar numbers are obtained by methods which do not depend on line
counting. For $2.5<z<4.5$, Press et al (1993), measuring $\tau_{\mathrm
eff}(z)$, derived $\gamma= 2.46\pm 0.37$.  Zuo \& Lu (1993), deriving
$D_A$ from spectra reconstituted from published absorption line lists,
find $\gamma= 2.87\pm 0.23$, but they considered a broken power law with
$\gamma = 2.82$ below and 5.07 above $z=3.11$ to give a better fit, which is
in agreement with an upturn at the highest redshifts.  The evolution of
the mean absorption $\tau_{\mathrm eff}$ or $D_A$ is indeed a more robust
measure of change, but its relation to the number of clouds is not
entirely straightforward as a model has to be relied on for the
distribution of absorbers, ${\cal F}(N,b,z)$,the functional form of
which is a priori unknown.  Although usually not taken into account, there
clearly is mutual dependence of $N$, $b$,
and $z$ in the form of clustering, and of
differential evolution in $N$ and $b$.  Moreover, the column density of
the lines dominating the absorption changes with redshift (at $z\sim 3$
lines with $\log N(HI)\sim 14$ contribute most), and so does the range
of column densities to which the measurements of $\tau_{\mathrm eff}$ are most
sensitive.

\medskip

To the surprise of many the first studies with HST of
low redshift ($z<1.5$) Ly$\alpha$ lines (Morris et al 1991; Bahcall et al 1991,1993;
Impey et al 1996) (see also the paragraph on the low
z Ly$\alpha$ forest below) have discovered more absorption systems than
expected from a naive extrapolation of any of the high z power law
exponents. Moreover, the low $\gamma=0.48\pm0.62$ valid from $z\sim 0$ 
to  $z\sim 1$ gives a $d{\cal N}/dz$ that is consistent with a
constant comoving density of objects. Accordingly, single power law
fits attempting to explain both high and low z absorbers  fail
to account for the rapid upturn around $z\sim 1$ (e.g., Impey et al 1996).

\medskip

{\small \noindent \it DIFFERENTIAL EVOLUTION AS A FUNCTION OF LINE
STRENGTH \ \ } A number of authors have made the case for differential
evolution, as a function of column density or equivalent width.  At the
high column density end, Lyman limit systems ($\log N > 17$) were found
to be consistent (Lanzetta 1988; Sargent et al 1989) with a
non-evolving population at least out to $z=2.5$. Therefore the lower column
density systems must be evolving faster, given the above $\gamma$
values.  Murdoch et al. (1986) and Giallongo (1991) noted a general
tendency toward  for slower evolution of $\gamma$ with increasing $W$ threshold,
which could provide
continuity between the large $\gamma$ for the general line population and the
non-evolving, optically thick Lyman limit regime.  However, other
studies claim the opposite trend.  Bechtold (1994) found that  $\gamma$
increased from $1.32\pm 0.24$ (for weaker lines (W$>$ 0.16 \AA ) to
$1.89\pm 0.28$ for strong lines ($W>$ 0.32 \AA ), Similar conclusions
are reached by Acharya \& Khare (1993). Bechtold's weakest lines
($0.16<W<$ 0.32 \AA ) are consistent with no evolution all, a result in
agreement with the Keck study by Kim et al (1997).  The latter group,
and Giallongo (1991) based their conclusions on high resolution data,
whereas the other papers cited are based on low resolution.  Blending
cannot be the whole explanation, since the present discrepancy persists
regardless of resolution.

\medskip

{\small
\noindent \it CONCLUSIONS \ \ } 
We may summarize the more secure results on number evolution
as follows: 
Going from $z=0$ to $z\sim 1$ there is no obvious change in the
comoving number of the clouds. Then between $1<z<2$ a steep rise sets in, which can be
reasonably described by a power law $(1+z)^{\gamma}$ with index $2<\gamma<3$. At
redshifts approaching $z\sim 4$, the upturn appears to steepen further. Thus a single
power law does not fit the curvature of the $d{\cal N}/dz$ vs. $z$
relation well. Differential evolution, with stronger lines evolving
less rapidly must exist to reconcile the large $\gamma$ for most of the
forest lines with the non-evolving population of Lyman limit absorbers.
Some studies suggest that the line density at the very low column
density range does not evolve much either, in which case the rapid
evolution inbetween is caused either by a genuine, rapidly changing
sub-population, or by biases in the analysis which we do not understand
properly.  In any case, the average optical depth is evolving
strongly with $\tau_{\mathrm eff}$ $\propto (1+z)^{\gamma +1}$ and
$\gamma$ around 2 -- 3 (for $z\simgt 3$).

\subsection{\it The Proximity Effect: Measuring the Ionizing UV Background}

The UV radiation from QSOs has been considered as the most natural
origin for the ionization of the intergalactic gas (Arons \& McCray
1969; Rees \& Setti 1970). The finite number density of QSOs 
suggests that there may be inhomogeneities in the ionization state of
the Ly$\alpha$ clouds near each QSO.  The term ``Proximity Effect''
(coined by Bajtlik et al 1988) refers to a relative lack
of Ly$\alpha$ absorption in the vicinity of the background QSO.  The
effect was first discussed by Weymann et al (1981), who
also suggested the currently accepted explanation of increased
ionization of the clouds by the nearby  QSO.  Carswell et al (1982)
realized that the general increase of the absorption line density $d{\cal
N}/dz$ with redshift was accompanied by a
simultaneous decrease of $d{\cal N}/dz$ in each individual QSO spectrum when approaching
the QSO's emission redshift.  Murdoch et al (1986)
calling it the inverse effect confirmed and clarified this result.
Tytler (1987b) referring to the effect as the anomaly, 
questioned the specific assumption that the reduced
absorption is caused by the QSO's radiation field. Hunstead et al
(1988), and Webb \& Larsen (1988) defended the earlier interpretation,
and in particular the claim that the effect is local to the neigborhood
of the QSO.

If the proximity effect is indeed caused by enhanced ionization 
measuring the intensity of the ionizing UV background becomes possible from observations
of the density of lines, $d{\cal N}/dz$, as a function of the
distance from the QSO.
Let us assume that in the vicinity of a QSO $d{\cal N}/dz$ is reduced, presumably owing to the
excess ionization of the gas.
With increasing distance from the emission redshift the QSO's ionizing flux
decreases until the UV background intensity begins to dominate the
ionization of the intergalactic gas. For example, at the point where the background intensity equals the QSO flux, $L_Q (4\pi r_L^2)^{-1}$ (known from photometry),
the neutral column density of a cloud should be lower by a factor of one half,
with a corresponding decrease in $d{\cal N}/dz$ for lines above a given detection
threshold.
In this way Carswell et al (1987) performed the first crude measurement of
the UV background radiation field, obtaining $J_{-21}$=3, where  $J$=$J_{-21}\times 10^{-21}$ erg
cm$^{-2}$ s$^{-1}$ Hz$^{-1}$ sr$^{-1}$ is the intensity at the Lyman limit, 912 \AA .
Bajtlik et al (1988) confirmed this result from a larger low resolution
sample of 19 QSOs, obtaining
$J_{-21}$=$1_{-0.7}^{+3.2}$. Their measurement procedure (adopted by most
later studies) consists of fitting
the number density of lines per unit redshift distance $X_{\gamma} = \int (1+z)^{\gamma} dz$
\begin{eqnarray}
\frac{d{\cal N}}{dX_{\gamma}} = \left(\frac{d{\cal N}}{dz}\right)_0 \left(1+\frac{L_Q}{16\pi^2 r_L^2 J}\right)^{1-\beta} 
\end{eqnarray}
as a function of the luminosity distance $r_L$, where the background
intensity $J$ is the quantity desired. 
The quantity $\beta$ is again the exponent of the power law distribution
of column densities.
Lu et al (1991)
with a larger sample get identical results. Bechtold (1994) in her even
larger dataset finds $J_{-21}$=3. The largest compilations of high
resolution data gave $J_{-21}$=$0.5\pm 0.1$ (Giallongo et al 1996), and
$J_{-21}$=$1^{+0.5}_{-0.3}$ (Cooke et al 1997).  None of
these studies has found evidence for a significant change with redshift
(for $1.6<z<4.1$). However,  Williger et al (1994) and Lu et al (1996) both
found lower values ($J_{-21}$=0.2)  just above $z\sim 4$, in
contrast to Savaglio et al's (1997) value ($J_{-21}$=0.7) for the
same redshift which is consistent with no evolution.  In any case, when 
going to lower redshifts
there appears to be a drastic decline in intensity below $z\sim 1.6$.
At  $<z>\sim 0.5$ Kulkarni \& Fall (1993) obtained
$J_{-21}$=$6\times10^{-3}$ from HST FOS data (Bahcall et al 1993).

\medskip

{\noindent\it \small SYSTEMATIC EFFECTS ? \ } Assuming that the
proximity effect really does measure the background intensity and not
some other environmental effect caused by the QSO (e.g., suppression of neutral gas
absorption in a high pressure environment)  we still
know this quantity only to within an order of magnitude, and the
uncertainty may even be larger than that. This is because, in addition
to the errors from line counting discussed in the previous section,
there are systematic uncertainties in the QSO flux.

Espey (1993) has quantified the overestimate in the local QSO flux
which arises from the well known systematic difference between the
``actual'' redshift of the QSO (as determined from narrow forbidden
lines) and the redshift of the (usually
blue-shifted) broad emission lines that define the ``QSO redshift'' used
normally for the proximity effect. The blueshift may typically
amount to $\sim 1500$ kms$^{-1}$. Downward corrections for  $J$
correspond to a factor of 2--3 in this case. These corrections have been
taken account in the work by Williger et al (1994), Lu et al (1996),
and Cooke et al (1997) but they would reduce for example
the Bechtold value by a factor three.

Other uncertainties more difficult to quantify (see the review by
Bechtold 1995) include QSO variability on the ionization time scale of
the gas (Bajtlik et al 1988), gravitational lensing amplification
of the apparent QSO luminosity, and uncertainties in the shape of the column density
distribution (Chernomordik \& Ozernoy 1993).
Gravitational clustering of clouds near the QSOs may cause an
excess number of clouds which again would lead to an 
overestimate of the background intensity by a factor up to three
(Loeb \& Eisenstein 1995).

The proximity effect measurements beg the question whether there are
enough QSOs to produce the ionizing background seen, or whether an
additional population of sources is needed. We may conclude as did Espey (see Espey 1993, and references
therein) that the contribution of known QSOs to the background
intensity agrees probably within the errors with the intensity from
this measurement.

A related debate on whether the QSOs can ionize the IGM has occupied an
even larger space in the literature. Unfortunately, we cannot do
justice to this extended discussion here, but refer to some of the
references in the section on the HeII Ly$\alpha$ forest below.

\subsection{\it Absorption Line Widths}

By measuring the line widths we may hope to gain insights into the
temperature and kinematics of Ly$\alpha$ clouds.  High resolution
spectra showed that many  low column density ($N_{HI} <
10^{15}$cm$^{-2}$) Ly$\alpha$ lines do indeed show widths ($b$ 
($=\sqrt{2}\sigma) \sim$ 10 - 45 kms$^{-1}$) that are consistent with
photoionization temperatures (Chaffee et al. 1983; Carswell et al.
1984; review by Carswell 1988), though some lines appear to be as wide as 100 kms$^{-1}$.
Median Doppler parameters are around $b_{med} \sim$ 30-35 kms$^{-1}$,
with a largely intrinsic scatter about the mean with standard deviation
$\sim$ 15 kms$^{-1}$ (Atwood et al 1985; Carswell et
al.  1991; Rauch et al. 1992).  The Doppler
parameters may decrease with increasing redshift.  Williger et al (1994)
have found an excess of lines with lower Doppler parameters $b\sim$
20kms$^{-1}$ at $z\approx$ 4.

\medskip

Occasionally, a correlation between Doppler parameter
and column density has been noted (Carswell et al.  1984, Hunstead et
al.  1988).  However, the reality of this effect has been subject to a
debate which culminated in the so-called ``b--N controversy'' (Hunstead
\& Pettini 1991; Webb \& Carswell 1991; Peacock 1991), when Pettini et
al. (1990)  suggested that (when looked
at with high enough resolution) Ly$\alpha$ lines had much lower Doppler
parameters (mostly $b$ $\simlt$ 22 kms$^{-1}$) than previously thought.
There also appeared to be a strong positive correlation between Doppler
parameter $b$ and column density $N$.  These results were not confirmed
by an analysis of another QSO spectrum with an identical observational
setup (Carswell et al.  1991).  The controversy was resolved when it was realized
that the
presence of noise in a spectrum can distort weak line profiles and lead
to underestimates of the average $b$ values of low column density
lines.  The problem is exacerbated by a detection bias
against weak broad lines, which  are more difficult to find against a
noisy continuum and tend to end up below the detection threshold.  The
combination of these effects accounts for both the presence of
spuriously low Doppler parameters and the apparent $b-N$ correlation
seen in these datasets (Rauch et al 1992, 1993).

\medskip

{\small \noindent \it RECENT KECK RESULTS \ \ } Data taken at similar
resolution but with much higher signal to noise ratio with the Keck
telescope's HIRES instrument have basically confirmed the earlier 4m
results.  Hu et al. (1995) found the Doppler parameter distribution at
$z\sim3$ to be well represented by a Gaussian with a mean of 28 kms$^{-1}$
and width $\sigma$ = 10 kms$^{-1}$, truncated below a cutoff $b_c$ = 20
kms$^{-1}$. With increasing redshift, there seems to be a genuine trend
to lower Doppler parameters. The finding by Williger et al (1994) of
evolution in $b$ appears confirmed:  Median Doppler parameters for
relatively strong lines (13.8$<$$\log$N(HI)$<$16.0) change from 41
kms$^{-1}$ ($<z>\sim$ 2.3; Kim et al. 1997) to 31 kms$^{-1}$ ($<z>
\sim$ 3.7; Lu et al 1996), with lower cutoffs dropping from 24 to 15
kms$^{-1}$ over the same redshift range.  The locus of the Pettini et
al (1990) narrow lines in the $b-N$ diagram  is virtually empty (Hu et
al 1995), as expected in data with such a high S/N ratio.  Kirkman \&
Tytler (1997) obtain similar results for the Doppler parameters in
their  Keck data at even better S/N ratios, but they question the significance
of the change with redshift, and find a lower, mean $b$ of 23
kms$^{-1}$ ($<z> \sim$ 2.7) with a lower cutoff $b_c$=14 kms$^{-1}$ at
$\log$N(HI)=12.5.  However, at $\log$N(HI)=13.8 their minimum $b$ at 19
kms$^{-1}$ is very close to the result of Kim et al. 1997, so the analyses
may well be consistent.  It is conceivable that the discrepancies at
lower column densities arise once more from the noise bias discussed
above which may affect any dataset as long as there continues to be a
supply of weaker and weaker lines crossing the detection threshold with
increasing S/N ratio.  The differences might lie in a different
understanding of what constitutes ``statistically acceptable fits'' or
``detectable lines''.  A dataset spanning a large redshift range, with
-- most importantly -- a homogeneous S/N ratio would be desirable.

\medskip

{\small \noindent \it THE TEMPERATURE OF THE IGM FROM LINE PROFILES
?\ \ } Though narrow lines ($b<15$ kms$^{-1}$) are apparently very rare
or even absent, this should not be interpreted as indicating a minimum
temperature of the Ly$\alpha$ absorbing gas. The issue is more complex;
in analogy with other astrophysical situations there are reasons for
which a correlation might be expected between the
temperature (or velocity dispersion) and the density (or column
density) of the gas.  Typical photoionization equilibrium temperatures
should be in excess of 30000 $K$ (e.g., Donahue \& Shull 1991), but
temperatures as low as 20000 $K$ can be attained through inverse
Compton cooling and a decrease of the ionizing spectrum at the HeII
edge (Giallongo \& Petitjean 1994). If photo-thermal
equilibrium is abandoned, adiabatic expansion cooling can lower the temperatures
further while maintaining high ionization, as suggested by Duncan et al (1991).  Currently favored theories of Ly$\alpha$
clouds that are the result of cold dark matter-based gravitational collapse do predict a
$b-N$ correlation with temperatures for low column density clouds even
below 10$^4$ $K$, as a consequence of  adiabatic expansion and inefficient
photoheating at low densities, while the larger column densities are
heated as a result of compression during collapse.  However, inspite of
the low temperatures the Doppler parameters of the weak lines are
predicted to be large because of bulk motion: Nature, in a random act
of unkindness, has endowed these cool clouds with a large size so that
the Hubble expansion
dominates the line broadening.  Yet, it may be worth trying to track down
the residual influence of the gas temperature on the line profiles.
The lower column density systems are at gas densities where the cooling
time (for processes other than expansion cooling) exceeds a Hubble
time. Therefore, the gas retains a memory of the temperature after reheating is complete (Miralda-Escud\'e \& Rees 1994; Meiksin 1994; Hui \& Gnedin
1997), and the process of reheating may have left a record in the
Doppler parameter distribution (Haehnelt \& Steinmetz 1997).
\medskip

\subsection{\it The Column Density Distribution}

In absorption line studies, the column density distribution function
(CDDF), i.e., the number of absorbers in a given column density bin
occupies a similarly, central place (and
provides similarly vague information) as the luminosity function in the
study of galaxies.  The observational determination of the CDDF relies
on a patchwork of techniques, owing to the large dynamic range of
observable HI absorption.  The CDDF can be measured relatively
unambigously from Voigt profile fitting for column densities between
N(HI) greater than a few times $10^{12}$ cm$^{-2}$, the detection limit
for typical Keck spectra, to about a few times $10^{14}$cm$^{-2}$,
where the linear part of the curve of growth ends.  The weakest lines
(N(HI) $\simlt 10^{13}$ cm$^{-2}$) are so numerous that they begin to overlap, 
requiring application of confusion corrections (Hu et
al. 1995).  Above $\sim 10^{14}$ cm$^{-2}$, once a line is saturated,
there is a certain degeneracy between a small change in apparent line
width and a large change in column density. Then the column densities
are relatively difficult to measure exactly by any means. The situation
is complicated further by the appearance of noticeable
multi-component structure in absorption systems with column densities
above 10$^{15}$ cm$^{-2}$ (e.g., Cowie et al 1995). Blending among
these components due to the large thermal width of the (already
saturated) Ly$\alpha$ can mimick large Doppler parameter/ high column
density lines. Simultaneous fits to the higher order Lyman lines (which
are less saturated owing to their lower oscillator strengths) can only
help to some degree. Eventually, for systems with column density 
N(HI)$\simgt$ 10$^{17}$ cm$^{-2}$ the discontinuity at the Lyman limit (LL)
can be observed (10$^{17.3}$cm$^{-2}$ corresponds to $\tau_{LL}\sim 1$) giving
again a relatively precise measure of the HI column (Lanzetta 1988;
Sargent et al 1989). From $N\simgt$ 10$^{18.5}$ on, the damping wings of
Ly$\alpha$ become detectable. The line width is now entirely a measure
of the column density and can again be read off by Voigt profile
fitting, or, in lower resolution data, directly from the equivalent
width (Wolfe et al 1986).

Carswell et al. (1984)
found that the number of absorbers ${\cal N}$ per unit HI column density interval can be
parametrized as
\begin{eqnarray}
\frac{d{\cal N}}{dN} \propto N^{-\beta}, \ \ \ \beta = 1.68 \ \ \ \ 13 < \log N < 15.
\end{eqnarray}
Tytler (1987a) has drawn attention to the remarkable fact that when results from higher column density
surveys are included a single power law with slope $\beta$ = 1.5 fits the whole
range of observable column densities well. Keck spectra (Hu et al. 1995, Lu et al. 1996,
Kirkman \& Tytler 1997, Kim et al. 1997) appear to show that the the power law extends
over ten orders of magnitude in column density from 10$^{12}$ to 10$^{22}$ cm$^{-2}$
(if the confusion correction made at the low column density end is justified).
A customary working definition of the CDDF (Tytler 1987a) is
\begin{eqnarray}
f(N) = \frac{{\cal N}}{\Delta N \sum_i \Delta_i X},
\end{eqnarray}
which gives the total number ${\cal N}$ of absorbers in HI column density bin $[N, N+\Delta N]$,
found over the total surveyed redshift distance $\sum_i\Delta_i X$, where $X$
has been defined earlier.

One of the more recent measurements (Hu et al 1995) gives
\begin{eqnarray}
f(N) = 4.9\times 10^7 N^{-1.46}, \ \ \ \ 12.3 < \log N < 14.5
\end{eqnarray}

Various authors (Bechtold 1987; Carswell et al 1987,
Petitjean et al. 1993; Meiksin \& Madau 1993; Giallongo et al 1993) have presented evidence
for departures from a single power law which seem to be borne out by
the new Keck data. A steepening for $\log N \simgt$ 14 explains why
individual high resolution spectra tend to yield $\beta \sim 1.7
-1.8$ for the regions ($\log N$ $\sim$ 13--15) for which they are most
sensitive (e.g. Carswell et al. 1984,1987; Atwood et al
1985; Rauch et al 1992). In the high column density range (beyond $\log
N$ $\sim$ 16) $f(N)$ flattens and damped systems are more abundant than
they should be judging from an extrapolation of the lower column
density power law. Weak evolution of $f(N)$ may occur in the sense
that this turnover moves slightly with redshift (Carswell et al 1987,
Kim et al. 1997) but the evidence is currently not overwhelming,
given that the dip in $f(N)$ occurs in a column density range, where the
determination of N(HI) is least certain (see above).

\subsection{\it Spatial Structure along the Line of Sight: Clustering and Voids}

Measurements of the two-point correlation function (TPCF) in velocity
space along the LOS, $\xi$($\Delta v$), led Sargent et al (1980) to
conclude that Ly$\alpha$ clouds are not clustered as strongly as
galaxies. Given the probability $\Delta p$ of finding a pair of
clouds with absorption cross section $\sigma$ and space density
$n_0(z)$, separated by a velocity interval $\Delta v$, $\xi$($\Delta
v$) is given by the following expression:
\begin{eqnarray}
\Delta p = n_0 \sigma \Delta v [1+\xi(\Delta v)],
\end{eqnarray}
where $\Delta v$ = $c\Delta z/(1+\overline{z})$ is the velocity
splitting in the rest frame at mean redshift $\overline{z}$.  No
correlation signal was found on scales of $\Delta v$ between 300 and 30000 kms$^{-1}$.
Clustering for small line pair splittings ($\Delta v \simlt$ 300 kms$^{-1}$)
would still be consistent with this result, given the limited
resolution, and the effects of blending caused by the large line
widths.  Based on Voigt profiles fits to high resolution data Webb
(1986) obtained the first evidence for the presence of a weak
clustering signal ($\xi$(100 kms$^{-1}$) $\approx 0.5$, at $z\sim$ 2.5) at small
separations.  This result has been confirmed by others (e.g., Muecket
\& Mueller 1987; Ostriker et al 1988).  It is hard to
detect the clustering at a high level of significance in any individual
QSO spectrum because of the short redshift path length, and both
detections and non-detections have been reported (Kulkarni et al 1996;
Rauch et al 1992). A variety of techniques seem to indicate, however,
that there really is weak clustering in the $z\sim 3$ forest on small scales. If
Ly$\alpha$ lines are considered as blends of components with
intrinsically narrower line widths the clustering amplitude could be
much higher (Rauch et al 1992).  In particular, the strong clustering
seen among metal absorption lines is largely invisible in Ly$\alpha$
because of blending between the saturated Ly$\alpha$ components
associated with the metals.  It is perhaps not that surprising
that a Ly$\alpha$ clustering amplitude, increasing with HI column
density was actually reported (Chernomordik 1994; Cristiani et
al 1995, 1997), and has been related to the clustering seen in metal
absorption lines (Cowie et al 1995; Fernandez-Soto et al 1996). Earlier, Crotts (1989),
by measuring the correlation in real space across the sky among systems in
multiple LOS had reported an increase of clustering with Ly$\alpha$
equivalent width.

\medskip

{\noindent\it \small STRUCTURE ON VARIOUS SCALES\ \ }The amplitude of
the TPCF is not the only tool for measuring structure in the forest.
In low resolution data, blends caused by clustering  show up as a
distortion in the equivalent width distribution of Ly$\alpha$ lines,
such that in the clustered case there are more large equivalent width
lines and fewer small ones than for a random distribution of lines in
velocity space (Barcons \& Webb 1991).  A number of other approaches,
mostly equivalent to the hierarchy of correlation functions, or parts
thereof (White 1979) may give a more robust clustering signal on small
scales by including higher order correlations.  Especially, the void
probability function  and, more generally, neighbor statistics
((Ostriker, Bajtlik \& Duncan 1988; Liu \& Jones 1990; Fang 1991;
Meiksin \& Bouchet 1995) have been used, invariably revealing the
non-Poissonian (clustered) nature of the distribution of clouds in
velocity space (Ostriker et al 1988; Bi et al
1989; Liu \& Jones 1990; Babul 1991).

Structure has been detected on many different scales, in addition to
the smale scale clustering described above: Fang (1991) used a
Kolmogorov-Smirnov test for nearest neighbor intervals to detect a
signal on scales 30-50 h$^{-1}$ Mpc (where h is the present day Hubble
constant in units of 100 kms$^{-1}$Mpc$^{-1}$). Mo et al (1992), from an analysis
of the extrema in the slope of the TPCF saw structure at 60 and 130
h$^{-1}$ Mpc. Meiksin \& Bouchet (1995) found an anti-correlation in
the TPCF around 3-6 h$^{-1}$ Mpc. Pando \& Fang (1996), applying the
wavelet transform, found clusters $\sim$ 20 h$^{-1}$ Mpc in size in the
Ly$\alpha$ forest. The physical interpretation of the various results
is not entirely obvious. The usefulness of the data for large scale structure
analyses has always been accepted at
face value, and it would certainly be entertaining to see whether there
are systematic effects in the data, perhaps causing some of the
structure. There are intrinsic scales in the spectra (like the
quasi-periodic change in S/N ratio caused by the sensitivity maxima of
the orders in an echelle spectrum) which are of similar magnitude
($\sim$ 5000 kms$^{-1}$, or $\sim$ 25 h$^{-1}$ Mpc comoving at z$\sim$
3) as some of the above detections.

Most of the clustering work is based on analyzing correlations
between distinct absorption lines. Including information about the
relative strength of the absorption as a function of velocity splitting
can improve the significance of any correlations, and give clues to the
mechanism causing the signal. Webb \& Barcons (1991) and Zuo (1992)
have suggested correlating equivalent widths, rather than just lines
above a detection threshold to search for inhomogeneities in the gas
pressure or ionizing flux along the line of sight. Fardal \& Shull
(1993), Press et al (1993), and Zuo \& Bond (1994) have extended
this approach to statistical models of the pixel intensity
correlations, a technique useful for disentangling line widths and
small scale clustering on overlapping scales.

\medskip

{\noindent\it \small VOIDS IN THE FOREST\ }
A specific discussion revolved around the question whether there are
voids in the Ly$\alpha$ forest, similar in comoving size to those seen
in the local galaxy distribution.  In principle verifying the existence of 
a void large enough to have a vanishing
probability of occuring by chance, if drawn from a Poissonian
distribution, is straightforward.  The probability function for a Poissonian gap of size
$\Delta z$ in a spectrum with absorption line density $d{\cal N}/dz$ is
simply 
\begin{eqnarray} P(\Delta z)=\exp\left[{-\left(\frac{d{\cal
N}}{dz}\right)\Delta z}\right].  
\end{eqnarray} 
Carswell \& Rees (1987)
concluded that voids with sizes like those in the local universe
($\sim$ 50 h$^{-1}$ Mpc (comoving)) cannot fill more than 5 \% of the
volume at $<z>\sim 3.2$. This result was confirmed by work by 
Duncan et al (1989), based on a larger dataset.
A similar conclusion was reached by Pierre et al (1988), who  found that
the Ly$\alpha$ absorbing gas cannot exhibit a void structure that is similar to
that of low redshift galaxies, without producing strong clustering
inconsistent with the observations.  

Nevertheless, individual large gaps have been found.  Crotts (1987),
discovered an 43 h$^{-1}$ Mpc gap towards Q0420-388.  This result was
variously contested and confirmed in a dispute about significance
levels (Ostriker et al 1988; Crotts 1989; Duncan et al 1989;  Bi et al 1991, Rauch et al 1992), fuelled,
among other things by the lack of a universally adopted definition of
the term ``void'' which takes into account that a void may be void of
lines only down to a certain detection threshold.  It turned out that
Crotts' gap, though not entirely empty, is a region of significantly
lower line density.  Dobrzycki \& Bechtold (1991) found another void of
size $\sim$ 32 h$^{-1}$ Mpc in the spectrum of Q0302-003, and Cristiani
et al (1995) discovered a significant pair of smaller voids towards
Q0055-269.  To summarize, the Ly$\alpha$ forest does show the occasional gap,
but the void structure apparent in the local galaxy population is not
present in Ly$\alpha$ absorbing gas.

The origin of the rare voids has most often been discussed in
connection with a local proximity effect: A foreground QSO near the LOS
to the more distant QSO produces a ``clearing'' by ionizing the
adjacent clouds seen in the LOS to the other object (Bajtlik et al 1988; Kovner \& Rees 1989).  No clean-cut evidence has been
found for this effect in the few studies done to date (Crotts 1989; M\o
ller \& Kjaergaard 1991), nor has it been possible to rule out its
existence (e.g., for the 0302-003 void, see Dobrzycki \& Bechtold 1991;
Fernandez-Soto et al 1995).  Several effects can complicate the
analysis.  Short of abandoning the idea of the proximity effect as an
excess of ionization caused by a nearby QSO, anisotropic
emission (beaming) by the QSO, and QSO variability can be invoked to explain the
non-detections and the discrepancies between the redshift positions of
QSO and candidate voids.  The number of free parameters in such models is
currently of the
same order as the number of voids observed, suggesting that studies of
individual voids will probably not be of great use for some time. With
a sufficiently large dataset, global searches for fluctuations in the
absorption pattern caused by an inhomogeneous radiation field (Zuo
1992; Fardal \& Shull 1993) may have a better chance of success.
However, Haardt \& Madau (1996) have recently pointed out that the
diffuse recombination radiation from Ly$\alpha$ clouds themselves (which can
provide on the order of 30\% of the ionization rate) can considerably reduce
the fluctuations.

\subsection{\it Spatial Structure Across the Sky: Multiple Lines of Sight}

The lack of two-dimensional information is one of the main shortcomings
of high redshift QSO spectroscopy. This problem makes it hard to understand the
geometry of the absorbers, and to disentangle positions in velocity and
real space. Through observations of common absorption systems in spatially separated
LOS (multiple images of gravitationally lensed QSOs, or QSO
pairs) the spatial dimension(s) across the sky can be restored to some
degree. Gravitationally lensed QSOs have maximum image separations up
to a few arcseconds, giving information on scales $< 100$ kpc.   In
contrast, QSO pairs rarely occur closer to each other than a few
arcminutes, which limits the scale that can be probed at high redshift
to larger than a few hundred kpc. The presence or absence of common
absorption in two LOS gives an indication of the coherence length of
the absorber. Obviously, common absorption systems must be at least as
large as the transverse separation between the beams to appear in both
LOS. If some of the systems are missing in one image,  a maximum
likelihood estimate based on the binomial probability distribution for
the fraction of ``hits'' and ``misses''  can be used to estimate the
mean extent of the absorbing cloud across the sky (McGill 1990).

\medskip

{\noindent\it \small GRAVITATIONALLY LENSED QSOS\ \ } The first gravitational lens
discovered, the z=1.39 QSO Q0957+561 (Walsh et al
1979) enabled Young et al (1981) to put a lower limit of $\sim7$ kpc on
the size of an intervening CIV absorption system common to both LOS.
Another notorious lensed QSO, 2345+007 A,B yielded lower limits to the
size of Ly$\alpha$ clouds of 1-11 $h^{-1}$ kpc (at z$\sim 2$) (Foltz et
al 1984; McGill 1990); the uncertainty comes from the unknown position
of the lensing object. This scale, though strictly an upper limit, came
to be considered as a typical ``size'' for several years, until the
observations of UM 673 (z=2.73; lens at z$\sim$ 0.5, LOS angular
separation 2.2 arcseconds) by Smette et al (1992)  showed two virtually
identical Ly$\alpha$ forests in the A and B images, with equivalent
width and velocity differences consistent with the measurement errors.
With LOS proper separations ranging up to 1h$^{-1}$ kpc for Ly$\alpha$
systems a 2--$\sigma$ lower limit of 12$h^{-1}$ kpc to the diameter of
the clouds (which were assumed to be spherical) was obtained.  
Observations of another bright lensed QSO (HE 1104-1805 A,B) by Smette
et al (1995a) produced even more stringent lower limits to cloud diameters 
of 25h$^{-1}$
kpc. These remarkable results seem to
indicate that the Ly$\alpha$ forest absorbers are not consistent
with the relatively small clouds envisaged in the pressure confined
model, nor are they consistent with the possibility that the absorption of a significant
fraction of the systems could arise in the virialized regions of
galaxies, or could exhibit the small scale variations typical of the interstellar
medium.

Unfortunately, very few lensed objects are suitable targets for
Ly$\alpha$ spectroscopy: At least two images must be bright enough to
be observed at high resolution; the image separation must exceed
typical seeing conditions ($>1.5$ arcseconds); the emission redshift
must be sufficient to shift the Ly$\alpha$ forest into the optical
wavelength range. The apparent lack of structure of the forest over
several kpc and the limited angular separation of lensed LOS make
observations of QSO pairs more suited for studying the large scale
structure of the Ly$\alpha$ forest proper. Lensed QSOs however, should
become increasingly useful for the study of metal absorption systems
and high redshift galaxies, a topic which is beyond the scope of this review
(see Smette 1995b).

\medskip

{\noindent\it \small QSO PAIRS\ \ } In view of the crowding in the high
redshift Ly$\alpha$ forest and the relatively large separations between
most QSO pairs, the cross-identification of individual absorption
systems in the two LOS can be difficult, and detections of coherent
absorption across the sky may be significant only in a statistical
sense. Sargent et al (1982) applied a cross-correlation
analysis to the spectra of the QSO pair near 1623+26 (Sramek \& Weedman
1978; see also Crotts (1989), who included two additional QSOs near
this pair). With transverse separations $\sim$ 1-2h$^{-1}$ Mpc the LOS
were well-suited for searching for the large coherent absorption
pattern predicted by some theories.  Oort (1981), for example, had
suggested that the Ly$\alpha$ forest was caused by intergalactic gas
distributed like low redshift ``superclusters''.  Sargent et al (1982) concluded that there is too little coherence between the
systems across the plane of the sky to agree with the large ``pancake''
structures envisaged by Oort.

However, evidence for very large structures seen in the much less
densely populated CIV forest suggests that some of the coherent
absorption in the Ly$\alpha$ forest is being missed because of the
difficulty of cross-identifying absorbers.  Therefore, examples of
gaseous structures with large coherence usually have come from high column
density systems. Shaver \& Robertson (1983) saw common absorption over
$\sim$ 380 $h^{-1}$ kpc at z$\sim 2$ in metal absorption systems (Q0307-195A,B).
Francis \& Hewett (1993) found coincident damped Ly$\alpha$ absorption
over $\sim$ 3h$^{-1}$ Mpc across the sky. Considering the firm lower limits
from gravitational lensing there was reason to expect large sizes for
the low column density forest as well, and such evidence was eventually
found:  Optical MMT data of the unusually close QSO pair Q1343+266A,B
(also known as  Q1343+264A,B; separation of 9.5 arcseconds) showed absorbers
extending over several hundred kpc at redshifts just below 2 (Bechtold
et al 1994; Dinshaw et al 1994).  Even larger sizes appear to occur at
somewhat lower redshift. DInshaw et al (1995) deduced a most probable diameter of
360 $h^{-1}$ kpc for
0.5 $< z <$ 0.9.from HST FOS spectra of the Q0107-25A,B
pair.  A later re-analysis points to a median
diameter of $\sim 0.5$h$^{-1}$ Mpc for Q1343+266 and and $\sim$ 1
$h^{-1}$ Mpc for the lower z Q0107-25A,B (Fang et al 1996).  The
velocity difference between lines in different LOS assumed to belong to
the same absorption system however, are small, with a intrinsic mean
difference of only $\sim$ 50 kms$^{-1}$.  Fang et al (1996) found a
trend for the size estimate to increase with separation between the
LOS, which is indicative of spatial coherence on a range of scales, the upper
end of which may not have been sampled yet.  Most of these studies
assumed spherical clouds, but from photoionization arguments 
spherical objects consistent with the measured sizes and column
densities would be so highly ionized as to easily overfill the universe
with baryons.  The simplest explanation, which allows for a denser and more
neutral gas while remaining consistent with the transverse sizes assumes that
typical Ly$\alpha$ clouds are flattened, with a thickness
on the order of $\sim$ 30 $h^{-1}$ kpc, for transverse sizes of 1 $h^{-1}$ Mpc
(Rauch \& Haehnelt 1995).

The large sizes, the small velocity and column density differences, the
possible absence of a unique size scale, the apparent flatness of
the clouds found at high and intermediate redshifts, and the general
absence of voids all argue against an origin of the typical (i.e., low
column density, $\log N\simlt 14$) Ly$\alpha$ forest line in 
potential wells of already formed galaxies.

\section{\large THE FIRST GENERATION OF MODELS }

Theoretical modelling of absorption systems can be
traced back to Spitzer's (1956) prediction  (expanded by Bahcall
\& Spitzer 1969) that normal galaxies have large gaseous halos giving rise to heavy element UV absorption lines.  Bahcall \& Salpeter (1965)
considered groups of galaxies, Arons (1972) suggested forming low mass
protogalaxies as the probable sites of Ly$\alpha$ absorption.  The
interpretation by Sargent et al 1980 of their observations of the
Ly$\alpha$ forest alerted researchers to differences between metal and
Ly$\alpha$ forest absorption systems, with the evidence  pointing away
from  galaxies, to distinct astronomical objects, intergalactic gas
clouds.

\subsection{\it Ly$\alpha$ Clouds confined by the Pressure of an Intercloud Medium}

If the Ly$\alpha$ absorbers correspond to overdense clumps of gas,
their persistence throughout of the history of the universe must be
either due to only a slow change in their properties, or to replenishment of the clouds
on a shorter time scale. An apparent lack of rapid
evolution in the properties of the forest (later shown to be a statistical fluke), and the short electron and
proton relaxation time scales and mean free paths appeared to justify
treating the clouds as ``self-contained entities in equilibrium''
(Sargent et al 1980).  A two
phase intergalactic medium was postulated, with the hot, tenuous
intercloud medium (ICM) in pressure equilibrium with the cooler and denser
Ly$\alpha$ clouds.  The standard version of the pressure confinement
model (Sargent et al 1980; Ostriker \& Ikeuchi 1983; Ikeuchi \&
Ostriker 1986) considers spherical, and, since gravity is ignored,
homogeneous clouds.   This model is self-consistent, but there are no very compelling physical reasons for
preferring pressure to gravitational confinement or to no confinement
at all, and the possibility of self-gravitating clouds as an
alternative was discussed  soon (Melott 1980; Black 1981).

Nevertheless, the pressure confinement model for Ly$\alpha$
clouds is appealing for several reasons: It combines the concept of a multiphase
structure of the intergalactic medium, familiar from the interstellar
medium, with the idea of separate entities, "clouds", in analogy to,
but different from galaxies.  A hot intercloud medium may have 
been a possible source of the X-ray background. In addition the explosion
scenario (see below) provided a theory of cloud formation. Finally, the model
made testable predictions, a rare but risky undertaking for astrophysical
theories, which paved the way to its eventual demise.

The basic properties of pressure confined clouds, as worked out in detail
by Sargent et al (1980), Ostriker \& Ikeuchi (1983), and  Ikeuchi \&
Ostriker (1986) can be summarized as follows:

The Ly$\alpha$ clouds are supposed to be in photoionization equilibrium
with an ionizing UV background. The gas is heated by photoionization,
and cools via thermal bremsstrahlung, Compton cooling, and the usual
recombination and collisional excitation processes.   The cloud
evolution consists of several phases, depending on the relative lengths
of cooling and expansion time scales.  The ICM is expanding
adiabatically by the cosmic expansion at all times because the high
degree of ionization does not allow for efficient photoionization
heating. The denser clouds embedded in the hot ICM start out in
isothermal expansion with a temperature fixed by thermal ionization
equilibrium ($T_c \sim 3\times 10^4 K$), until the density $n_c\propto
P_{\mathrm ICM}/T_c\propto (1+z)^5$ has dropped sufficiently that photoheating cannot
compensate for the work of expansion any longer, and the clouds begin to
cool and to expand less rapidly. The sound speed drops even faster so
ultimately pressure equilibrium with the ICM ceases and the clouds
enter free expansion.  The available range of cloud masses is
constraint by the requirements that the clouds must be small enough not
to be Jeans-unstable, but large enough not to be evaporated rapidly when
heated by thermal conduction from the ambient ICM (Sargent et al 1980,
Ostriker \& Ikeuchi 1983). Clouds formed at $z\sim 6$ would survive
down to accessible redshifts ($\sim 4$) only if their masses range
between 10$^5 < M_c < $10$^{10}$ $M_{\odot}$.

\medskip

{\noindent\it \small THE ORIGIN OF LYMAN ALPHA CLOUDS FROM COSMIC SHOCKS \ \ }
The explosion scenario of structure formation (Schwarz et al 1975; Ostriker \& Cowie 1981) provided a non-gravitational origin
for the pressure confined Ly$\alpha$ clouds. Large scale explosions
from galaxies (e.g., from starbursts) and QSOs may have driven shock waves
into the intergalactic medium. These events may also have provided the
energy for collisional re-ionization and heating of the ICM, since
photoionization cannot produce temperatures high enough to maintain
pressure confinement (Ikeuchi \& Ostriker 1986).  This way a two-phase
medium of hot ``cavities'' enclosed by a system of cooler shells could
have arisen (Ozernoy \& Chernomordik 1976). Ly$\alpha$ absorption is
caused by the fragmenting shells (Chernomordik \& Ozernoy 1983;
Vishniac et al 1985). Among the observational
consequences may be pairs of Ly$\alpha$ lines that occur wherever the
line of sight intersects  an expanding, spherical shell. It has been
argued that such pairs have been seen (Chernomordik 1988, and refs.
therein); this may be the case in individual systems, but it is
difficult to prove in a statistical sense because the two-point correlation
does not show the expected signal at the relevant velocity scale,
$\sim$ 100 kms$^{-1}$ (Rauch et al. 1992).  A potential problem for this model is
implied by the observed lack of a correlation with galaxies.  If Ly$\alpha$
clouds are shells expelled by galaxies, the absorbers should be
clustered in a manner similar to that of galaxies (Vishniac \& Bust 1987).  Neither the
auto-correlation function among absorbers nor the cross-correlation
with galaxies show a signal of the requisite strength (Barcons \& Webb
1990). 

A similar pattern of shell formation ensues if
QSOs, in the initial event of reionization, surround themselves with
Str\"omgren spheres (Arons \& McCray 1969, Shapiro \& Giroux 1987),
which can lead to shocked shells of gas at the boundary of the HII
regions. The shells fragment as in the explosion scenario and may be visible as Ly$\alpha$
absorbers (Madau \& Meiksin 1991).

\medskip

{\noindent\it \small THE ELUSIVE INTERCLOUD MEDIUM \ \ } The need for a
confining intercloud medium has led to a number of searches for a
residual absorption trough between the absorption lines, caused by the
HI in the intercloud space.  These measurements came to be referred to
in the literature as the Gunn-Peterson (GP) test, though 
Gunn \& Peterson's (1965) original result of a 40\% average absorption was a
detection of the unresolved Ly$\alpha$ forest as a whole. Values for
the residual (i.e., intercloud, or diffuse, as opposed to line) GP
effect require (a) a precise knowledge of the unabsorbed QSO continuum
level, and (b) either subtracting the contribution from Ly$\alpha$
``lines'', or the use of line free regions (whatever that may be) for
the estimate. Steidel \& Sargent (1987b) measured the total flux
decrement of a sample of 8 QSOs against continua extrapolated from the
regions redward of Ly$\alpha$ emission.  After subtracting a
model population of discrete Ly$\alpha$ lines they obtained a
residual $\tau_{GP}<0.02\pm 0.03$ ($<z>\sim 2.67$), i.e., a null
result.  Giallongo et al (1992), and Giallongo et al
(1994) have compared apparently line free regions  with an
extrapolated continuum, and they found  $\tau_{GP}<0.013\pm
0.026$($<z>\sim3$) and ($\tau_{GP}<0.02\pm 0.03$($<z>\sim 4.3$),
respectively. Given the line crowding at high redshift, the small error
bars of the $z>4$ work betray a certain degree of optimism.

The clean-cut decomposition into line and continuum absorption makes theoretical
sense for pressure confined clouds, but observationally we can never be
sure whether there is a flat continuum trough from diffuse gas, or
whether there are many weak lines blended together. 
Jenkins \& Ostriker (1991), and Webb et al (1992), acknowledging this
problem, attempted to model the pixel intensity distribution with
continuum absorption and variable contributions from discrete lines.
Their results show that $\tau_{GP}$ can be produced both ways,
by blending of weak lines below the detection threshold, or by a constant
pedestal of absorption.

In any case the weakness or non-detection of a residual GP trough puts an upper
limit on the density of the ICM. A lower  limit on the ICM pressure
($\propto n_{ICM}T_{ICM}$) can be derived from the absorption line
width (which gives an upper limit on the radius of the expanding cloud
(Ostriker \& Ikeuchi 1983).  The condition that the cloud must be large
enough not to evaporate gives an upper limit on the pressure.  Another
independent upper limit on the pressure of the ICM comes from the lack
of inverse Compton distortions in the spectrum of the cosmic microwave
background (CMB) (Barcons et al 1991). This result rules out the
intergalactic medium as the source of the hard X-ray background. 
It also may spell trouble for the explosion model of galaxy formation, which is the origin of the
two-phase IGM in the current picture) and of apparently non-existing structure in the CMB. When all the limits are combined all the limits only a relatively small corner of allowed ($n,T$) parameter
space remains for the intercloud medium.  

\medskip

{\noindent\it \small PROBLEMS WITH THE COLUMN DENSITY DISTRIBUTION\ \ } For
pressure confined clouds the large range of neutral hydrogen column
densities observed must correspond to a range in the parameter
combination
\begin{eqnarray}
N(HI) \propto M_c^{1/3}T_c^{-29/12}J^{-1}P^{-5/3}.
\end{eqnarray}
To reproduce only the low column density systems between $13<\log N(HI)<
16$ the mass has to vary by 9 orders of magnitude, or the radiation
field by 3 orders, or the pressure by a factor of 63. To ensure cloud
survival the mass range is limited to less than 4 dex (see above), and the
temperature is constant; therefore, we may need to invoke pressure inhomogeneities (Baron et al 1989). However, Webb \& Barcons (1991),
looking for pressure related spatial correlations among  the equivalent
widths of Ly$\alpha$ forest lines excluded pressure fluctuations $\Delta
P/P > 14\% $ at the 2$\sigma$ level, and a similar limit must hold for
the radiation field $J$.  Extremely flattened clouds would help somewhat in
that they  would increase the column density range, allowing a wider range of path
lengths through the clouds (Barcons \& Fabian 1987), but that may introduce other
problems. Williger \& Babul (1992) taking these constraints
into account investigated pressure confined clouds with detailed
hydrodynamical simulations and found that the small mass range leads
not only to a failure in producing the column density range but also to
a faster drop in the number of clouds with redshift, than observed.

\medskip 

To summarize, the pure pressure confinement model is unlikely to explain
the Ly$\alpha$ forest as a whole, though it is clear that some LOS must
go through sites where gas is locally confined by external hydrostatic or ram
pressure.  Low redshift gaseous galactic halos, the likely hosts of the
dense Lyman limit absorbing clouds, may be such environments.
Formed by local instabilities the dense clouds may be in pressure
equilibrium with a hot gas phase at the virial temperature of the
halo (Mo \& Miralda-Escud\'e 1996, and references therein).

\subsection{\it Gravitational Confinement: Selfgravity}

Self-gravitating baryonic clouds were suggested by Melott (1980) as an
alternative to the pressure confinement model. Black (1981)
investigated in detail the physical structure of such objects.
clouds. He found that (quasi-)stable clouds with properties consistent
with the observations have to be extended ($\sim$1 Mpc), and must either be truncated
by an external medium, or be large enough to overlap, providing their
own boundary pressure. In this model the appearance of the
intergalactic medium as a forest of lines is more due to the strong internal
gradients of the neutral gas density, than to a sharp transition
between separate entities. The density of the intergalactic medium along the LOS
undulates, and there is no real difference between an intercloud medium
and the clouds. The huge sizes would also have been able to reconcile a
larger mass density of the intergalactic medium with the observed cloud
parameters, whereas pressure confined clouds would contain only a small 
fraction of all baryons.  The
model met with scepticism because the large sizes appeared
to contradict the scant observational evidence.  However it may also have other problems, such as reproducing the
column density distribution (Petitjean et al 1993a).

\subsection{\it Gravitational Confinement: Clouds Dominated by Hot Dark Matter Gravity}

The advent of ab initio theories of gravitational structure
formation  made it possible to place the Ly$\alpha$ forest in a larger
frame, and investigate its relation to galaxy formation. In
principle, the number density, sizes and physical parameters  of the
absorbers can be predicted as a consequence of cosmological models,
though, until recently this has been wishful thinking.

\medskip

{\noindent\it\small ZELDOVICH PANCAKES\ } The first such theory to
explicitly address the Ly$\alpha$ forest phenomenon was the hot dark
matter (HDM) model.  The formation of adiabatic pancakes suggested by
Zeldovich (1970) is expected to produce a primordial gas phase with the
right properties for detection in HI absorption (Doroshkevich \&
Shandarin 1977). Among the interesting consequences of this theory are
the large (in fact: too large) sizes of the pancakes. Even after
fragmentation, coherent absorption should be extending over Mpcs
across the sky (Doroshkevich \& Muecket 1985) thus explaining the large
transverse sizes seen later in QSO pair studies.  In a prescient paper,
Oort (1981), at the time referring to HDM pancakes, suggested
identifying Ly$\alpha$ absorbers with collapsed but uncondensed gas in
the ``superclusters'' of galaxies known from low redshift. He
calculated the gravitational scale height of a sheet of gas, and noted
the similarity of the mean free path between Ly$\alpha$ clouds and the
distances between superclusters.  The very large transverse sizes
expected ($\sim$ 20 Mpc, at $z\sim 2$) were not confirmed, however, by
the QSO pair study of Sargent et al (1982), so such objects
cannot be the rule.

The underlying HDM structure formation scenario has become somewhat
unpopular, but the physical idea of Ly$\alpha$ absorbers as flattened
pancakes survives into the currently favored, CDM based picture (see
below). However, the CDM pancakes (or sheets) are more than an order of
magnitude smaller than the Zeldovich ones, and they form late, after
denser structures like knots and filaments are already in place
(Cen et al 1994; Bond et al 1996).

\subsection{\it Gravitational Confinement: Cold Dark Matter Minihalos}

The properties of gas clouds under the influence of the gravitational
field of dark matter have been investigated by Umemura \& Ikeuchi
(1985), and, more specifically in terms of the ``minihalo'' model by
Rees (1986) and Ikeuchi (1986). In this picture, Ly$\alpha$ clouds are
a natural by-product of the CDM structure formation scenario.
Photoionized gas settles in in the potential well of an isothermal CDM
halo. The gas is stably confined, if the potential is sufficiently shallow to avoid
gravitational collapse but deep enough to prevent the warm gas from
escaping. The CDM minihalos are more compact than the self-gravitating
baryonic clouds of Black (1981) because of the larger dark matter
gravity. The detailed structure of the halo depends on the relative
spatial distribution of baryons and CDM.  The models can be
parametrized by the intensity of the radiation field $J$, the central
overdensity $\delta$($r$=0), and the ratio of baryonic to dark matter
(Ikeuchi et al 1988).  The minihalo model has the
attractive feature of providing a natural explanation for the overall
shape of the observed column density distribution function (CDDF).
The large observed dynamic range in column density reflects the strong
density variations as a function of impact parameter, rather than
a range in cloud properties.  For the general case where the baryon
distribution is an isothermal sphere ($n_b \propto r^{-2}$), the HI
density in the highly ionized region of the minihalo drops like
$n_{HI}$ $\propto r^{-4}$ with radius $r$, and the resulting column
density distribution seen by random lines of sight through a population
of such halos obeys $d{\cal N}/dN_{HI} \propto N_{HI}^{-1.5}$ (Rees
1988; Milgrom 1988).  The largest column densities, including damped
Ly$\alpha$ systems, are caused by the neutral cores in the shielded
centers of the clouds (Murakami \& Ikeuchi 1990).  Thus  minihalos can
produce a column density power law over almost nine decades, providing
a physical basis for Tytler's (1987a) suggestion of a common origin for all
QSO absorbers.  Evolution with redshift is caused by a number of processes
(Rees 1986):  When loosing pressure support as the ionizing flux decreases,
gas may settle deeper into the potential well, thus
reducing the geometric absorption cross-section. once stars are forming the UV
flux may rise again, and  stellar winds may blow out the gas (Babul \& Rees 1992),
thus increasing the absorption cross-section.  Halos are produced by a
continuing turn-around of density peaks, and are destroyed by merging.

A non-stationary version of the minihalo model was
studied by Bond et al (1988), who examined the hydrodynamics
of a collapsing spherical top-hat perturbation. If the accreting
baryonic component escapes gravitational collapse (and subsequent star
formation) it may reexpand under the influence of photoionization
heating and even recollapse after the UV intensity has ebbed (see also
Murakami \& Ikeuchi 1993).

In order to investigate the relative importance of various confinement mechanisms
Petitjean et al (1993a) have studied a hybrid model, of  spherical
clouds with or without dark matter, bounded by an external pressure.
These models show that pure baryonic clouds (as discussed by Black
1981), in order to be stable against the outer pressure, tend to overproduce
high column density systems.  Many of the observed features of the
Ly$\alpha$ forest clouds can be explained with a single type of
minihalo, but to match the fine structure of the column
density distribution function halos may be required to exhibit a
range of central densities (Murakami
\&  Ikeuchi 1990; Petitjean et al 1993b).  Two separate populations,
one for low column density Ly$\alpha$ clouds, and one for the higher
column density metal systems, give a better fit to the CDDF (Petitjean
1993b).  

Charlton et al (1993,1994) have studied gravitational
condensations with a different geometry, modelling Ly$\alpha$ clouds as
equilibrium slab models that are subject to the pull of CDM gravity and to an
external pressure. It was found that the change of slope in the column
density distribution function (near $\log$ N(HI) $\sim 15$) can be explained by
a transition between pressure and gravitational confinement, in the
sense that at higher column densities gravity takes over and imposes a
steeper dependence of the neutral column density with total column
density.

\medskip

{\noindent\it \small NON-EQUILIBRIUM AND OTHER EFFECTS POINTING
BEYOND THE SIMPLE HALO MODEL\ \ }

With the adoption of the CDM based models researchers could avail
themselves of the analytical apparatus developed to describe the dark
matter distribution, especially the important concept of ``halos'' (Babul
1990; Mo et al 1993).  Of course, there is a limit
to the degree of realism with which a counting scheme for dark matter
condensations or spherical collapse can describe the observed properties of gas
clouds, and even the notion of distinct ``objects'', dear to traditional
astronomy, may fail. Hydrodynamic simulations (see below) show that in a hierarchical
universe  at intermediate redshift ($\sim 2$) most baryonic matter may
not have settled in spherical, or rotationally supported virialized
objects, as suggested by the word ``halo''. Virial radii of objects
capable of stably confining HI clouds are $\sim 10$ kpc (Rees 1986).
The coherence lengths of Ly$\alpha$ systems from gravitational lensing
constraints (Smette et al 1992,1995) are much larger, implying that
only few LOS ever hit the virialized region.  Thus, minihalos may 
not only be embedded in regions of uncollapsed gas, they may still
be  accreting matter at the epoch where we observe the Ly$\alpha$ forest.
Thus it makes sense to look for signs of non-equilibrium, especially
departures from thermal line line profile caused by the bulk motion of
the infalling gas (Miralda-Escud\'e \& Rees 1993). Meiksin (1994) has
traced the formation and internal structure of minihalos and slabs with
hydrodynamical simulations to search for such observable
non-equilibrium effects. For a slab or pancake geometry, noticeable
deviations from Voigt profiles are predicted, but they would be hard to detect for
spherical clouds.

\section{\large THE COSMOLOGICAL SIGNIFICANCE OF THE LYMAN ALPHA FOREST}

Over the past few years semi-analytical work and in particular
hydrodynamic simulations of hierarchical structure formation have
gradually led to a minor Copernican shift in our perception of the
material content of the high redshift universe.  If the
inferences (discussed below) are interpreted correctly the intergalactic medium is the
main repository of baryons down to redshifts at least as low as $z\sim
2$.  If so, then high redshift galaxies -- in absorption line parlance ``Lyman
limit'' or ``damped Ly$\alpha$ systems'' -- are mere tracers of the
matter distribution.  The simulations show that the Ly$\alpha$ forest
is produced by a hierarchy of gaseous structures, with typical shapes
changing from sheets through filaments  to spherical galactic gaseous
halos, as the column density increases.  Perhaps most importantly,
Ly$\alpha$ forest lines closely reflect gravitationally induced density
fluctuations in the general matter distribution. Given the relatively
simple physics of this baryonic reservoir and the enormous sensitivity
of the observations Ly$\alpha$ forest spectra should make excellent and
largely unbiased probes of structure formation at high redshift.

\medskip 

{\noindent\it\small THE LARGE BARYON CONTENT\ \  } The fraction of
matter incorporated into galaxies or still left in the intergalactic
medium depends strongly on the structure formation model.  To calculate
the baryon content of Ly$\alpha$ clouds we need to know the ionization
correction, as most of the gas is highly ionized.  For a given ionizing
radiation field the degree of ionization depends on the density and
thus, for a given observed column density, on the spatial extent of the
gas. Deriving the mass content then requires fixing the size (or
scale height) of the clouds either from measurement, or from theoretical
prejudices.  For example, the fraction of mass required to cause the
observed amount of absorption can be quite large for gravitationally
confined, extended, baryonic clouds (Black 1981). In contrast,  the
small baryon content expected if the Ly$\alpha$ were caused by pressure
confined clouds (Sargent et al 1980) is largely a result of the small
cloud sizes adopted.  By using a suitable choice of parameters, the Ly$\alpha$
forest can be made to contain anything from a negligible fraction up to
virtually all of the baryons, and still be consistent with the
observations (Meiksin \& Madau 1993).  Specifically for the CDM
minihalo model, Petitjean et al (1993b) found that the Ly$\alpha$ forest
clouds had to contain most of the baryons at redshift 2-3, in order to
match the observed column density distribution function.  This is in
agreement with Shapiro et al (1994), who found that in a CDM
model the fraction of baryons not yet collapsed into galaxies should be on the
order of 50--90\% .  Independent of the cosmological model, the large
transverse sizes of Ly$\alpha$ absorbers measured from QSO pairs give
another, indirect indication that the baryon density in Ly$\alpha$
clouds must be large, or the absorbers must be extremely
flattened (Rauch \& Haehnelt 1995).
  
\medskip

{\noindent\it\small A FOREST OF LINES, OR A FLUCTUATING GUNN-PETERSON
EFFECT ?\ \ } 

Under the influence of gravity the intergalactic medium becomes clumpy
and acquires peculiar motions, and so the Ly$\alpha$ (or GP) optical
depth should vary even at the lowest column densities (Black 1981;
McGill 1990; Bi et al 1992; Miralda-Escud\'e \& Rees 1993;
Reisenegger \& Miralda-Escud\'e 1995). In a CDM dominated structure formation
scenario the accumulation of matter in
overdense regions reduces the optical depth for Ly$\alpha$ absorption
considerably below the average in most of the volume of the universe,
leading to what has been called the fluctuating Gunn-Peterson
phenomenon.  Traditional searches for the GP effect that try to measure
the amount of matter between the absorption lines are no longer very
meaningful as they are merely detecting absorption from matter leftover
in the most underdense regions. If this is not taken into account the
amount of ionizing radiation necessary to keep the neutral hydrogen GP absorption
below current detection limits can easily be overestimated.

As another consequence, the distinction between the low column density
Ly$\alpha$ forest ``lines'', and the GP ``trough'', becomes
somewhat artificial.  Bi and collaborators (Bi et al 1992;
Bi 1993; Bi \& Davidsen 1997) have shown that the optical depth
fluctuations corresponding to the linear regime of gravitational
collapse in the intergalactic medium can give a remarkably realistic
representation of the Ly$\alpha$ forest (ignoring the higher column
density lines, which are produced from non-linear structures, e.g.,
minihalo type objects). Their semi-analytical work is based on a
log-normal density fluctuation field. For low densities where
dissipation is not important the collapse of dark matter and baryons
differs mainly by the presence of the gas pressure which effectively
smooths the baryons distribution on scales below the Jeans length. Bi
et al treated the pressure as a modification to the power spectrum of the
baryon density contrast $\delta_b$, suppressing power on scales below
the Jeans length:  \begin{eqnarray}
\delta_b(k)=\frac{\delta_{DM}(k)}{1+k^2(\lambda_J/2\pi)^2}
\end{eqnarray} where $\lambda_J$ is the Jeans length, $k$ the
wavenumber, and $\delta_{DM}$ the dark matter overdensity.  This method
can elucidate many of the basic features of low column density
Ly$\alpha$ clouds.  The schematic treatment of the equation of state
and the lack of inclusion of shock heating limit the approach, however,
to overdensities of $\delta<5$, where gas physics beyond the Jeans
criterion is not very important.

\subsection{\it Hydrodynamic Simulations of the Ly$\alpha$ Forest}

{\noindent\it\small NUMERICAL APPROACHES\ \ }
From the early 1990s on hydrodynamic cosmological simulations  became
sufficiently realistic to be able to quantitatively predict the
physical properties of the intergalactic medium and the high redshift
Ly$\alpha$ forest from the initial conditions of a given structure
formation model (e.g., Cen \& Ostriker 1993). A Ly$\alpha$ forest
spectrum is completely specifed by the Hubble constant, gas density,
temperature, peculiar velocity, and neutral fraction along the LOS.
By predicting these quantities for artificial QSO LOS through simulated
slices of the universe it becomes possible to examine the
correspondence between Ly$\alpha$ forest absorbers and the physical
properties of the underlying gaseous structures.  This approach was
first taken by Cen et al 1994 (see also Miralda-Escud\'e et al 1996)
using an Eulerian hydro-simulation of a $\Lambda$CDM model.  Since then
a range of other numerical techniques have been applied to different
cosmological models.  The basic properties of the Ly$\alpha$ forest
turn out to be only weakly dependent on the cosmological model, and
similar answers have been obtained with a variety of approaches:
Petitjean et al (1995) grafted the baryons onto a COBE
normalized cold dark matter distribution from an DM particle mesh
simulation, using an analytic prescription to track the thermal history
of the gas. A standard CDM model has been studied by Zhang et al (1995, 1997) with an Eulerian code, and by Hernquist et al (1996),
with  a Lagrangian, Smoothed Particle Hydrodynamics (SPH) technique.  As
a crude general rule of thumb the Eulerian codes are capable of higher
resolution for the void regions producing  the lowest column density
Ly$\alpha$ forest, whereas the Lagrangian codes are superior for
regions like minihalos or galaxies where a larger dynamic range is
required. Thus the use of SPH codes has been extended to study damped
Ly$\alpha$ systems (Katz et al 1996) and metal absorption systems
(Haehnelt et al 1996a).  Hybrid schemes (e.g., Wadsley \&
Bond 1997) can be tailored to capture the influence of both, large
scale (long wavelength) gravitational effects and the small scale gas dynamics, on the
formation of Ly$\alpha$ absorbers.

\medskip

{\noindent\it\small  THE NATURE OF LYMAN ALPHA ABSORBERS \ \ } Inspite
of some quantitative differences a generic picture  of the Ly$\alpha$
forest has emerged from these studies:    Low column density systems
($\log N(HI) \simlt 14$) are associated with sheet-like structures, not
unlike small versions (length scale $\sim$ a few hundred kpc to 1 Mpc
proper) of Zeldovich pancakes. Gas accretes through weak shocks
(creating a double humped temperature profile), and settles in a dense,
central cooling layer, presumably to form stars. At the lowest column
densities gas remains unshocked and just bounces back because of the
hydrostatic pressure. The gas is partly confined by gravity and partly
by ram-pressure.  Higher column density clouds arise in more
filamentary structures, with column density contours of $\log N(HI)
\sim 14$ extending continously and with relatively constant thickness
($\sim 40 - 100$ kpc proper) over Mpc distances. With increasing column
density the absorber geometry becomes rounder; column density contours
at $\log N(HI) \simgt 16$ invariably are spherical, entering the regime
where the absorbers more closely correspond to minihalos; there the
enclosed gas column is high enough to make the absorption system appear
as a Lyman limit or damped Ly$\alpha$ system.  Figure 2
shows the spatial appearance of the Ly$\alpha$ absorbers.  The
visual appearance of the low column density, sheetlike-filamentary structure
has been aptly described as a ``Cosmic Web'' (Bond \& Wadsley
1997).  Looking at the higher column density, optically thick gas on
scales of several Mpcs one gets a somewhat different impression of
chains of mini- or larger halos, lining up like pearls on a string,
quite similar to the structure seen in N-body simulations of the dark
matter distribution.  Confirming earlier analytical work, a large
fraction of all baryons (80 - 90 \%) resides in the low column density
Ly$\alpha$ forest, mostly in the column density range $14 < \log N(HI)
< 15.5$ (Miralda-Escud\'e et al 1996).

A glance at a typical density-temperature diagram (Figure 3) for
random lines of sight through one of the SPH simulations (Haehnelt et al  1996b) reveals significant departures  from thermal
photoionization equilibrium for all but the highest density gas ($
n_H<10^{-3}$cm$^{-2}$).  The temperature density relation is generally
steeper than the equilibrium curve, because the lower density gas cools
by expansion, while the gas in the density range $n_{H}\sim$ a
few times $10^{-5}$ -- $10^{-3}$ cm$^{-3}$ is heated by
adiabatic compression or shock heating. Temperatures below $10^4 K$
occur in voids where the expansion velocity is largest. 

The gas is
still being accreted at the epoch of observations ($z\sim3$). Nevertheless,
the lower column
density flattened gas structures expand in proper coordinates because
the gravitational pull decreases together with the dark matter surface density,
as the universe expands.  Many of the weaker absorption lines arise in
low density, relatively extended regions, which expand with a
substantial fraction of the Hubble velocity. The expansion, and the low
temperatures due to the low density and the adiabatic cooling in voids ensure that at column
densities ($\log N(HI) \simlt 13$) bulk motion becomes the dominant source of
line broadening (Miralda-Escud\'e et al 1996; Weinberg et al 1997).

\medskip

{\it\noindent\small MATCHING THE OBSERVATIONS\ \ } The simulations have
been quite successful in matching the overall observed properties of
the absorption systems, and the agreement ranges from the acceptable to
the amazing. The shape of the column density distribution and the
Doppler parameter distribution are reasonably well reproduced by the
simulations.  (Cen et al 1994; Zhang et al 1995,1997; Hernquist et al
1996; Miralda-Escud\'e et al 1996). Although the approximate range of
Doppler parameters is hard to miss (with photoionization being the
great equalizer), subtle effects can raise or lower the mean line width
by $\sim 30\%$ and change the shape of the Doppler parameter
distribution. There may be some discrepancy for the Doppler parameters
between different simulations (Zhang et al 1997, Dav\'e et al 1997,
Miralda-Escud\'e et al 1996) but it is not yet clear whether this is
due to different types of data analysis, different assumptions about
the process of reionization, or limited numerical resolution.  A
departure from Voigt profile shapes, especially the broad wings of weak
lines signifying bulk motion broadening in sheets, is seen in the
simulations (Cen et al 1994) and appears to be present in real high
resolution spectra (Rauch 1996).  The large transverse sizes of the absorbers
seen against background 
QSO pairs and lensed QSOs are readily explained by the coherence length
of the sheets and filaments (Miralda-Escud\'e et al 1996; Charlton et
al. 1997; Cen \& Simcoe 1997).  The weak clustering amplitude appears
to be in agreement with the observations. The histogram of residual
fluxes in the Ly$\alpha$ forest is reproduced very well by the models
(Rauch et al 1997). Conversely, we may take this as observational
evidence in favor of some sort of hierarchical structure formation 
model.

The evolution of the Ly$\alpha$ forest with time at high redshift is
mainly driven by the Hubble expansion and the resulting increase in the
mean ionization of the gas, and to a lesser degree by the gas streaming
along the filaments (Miralda-Escud\'e et al 1996).  Muecket et al
(1996), from their simulation, find that the number of absorbers per
redshift is given by a broken power law, with $\gamma\sim2.6$
$(1.5<z<3)$ and $\gamma\sim0.6$ $(0<z<1.5)$, ($\log N(HI) > 14$)
(Riediger et al 1998), a remarkable agreement with the
observed data. The break in the power law  can be understood as a
change with time in the dimensionality of the structures dominating the
absorption.  The sheetlike absorbers dominating the high redshift
Ly$\alpha$ forest are expanding with time and are dropping below the
detection threshold first because of their low column density, leaving
the absorption from the less rapidly evolving gas distribution in the
filaments and knots to dominate. There the column density also
decreases, but since the original column was higher, the filaments
remain visible for longer. Continuing infall also contributes to the
increasing prominence of the more compact structures.

\subsection{The Ly$\alpha$ Forest as a Cosmological Laboratory} 

The first generation of simulations was largely aimed at establishing
the physical properties of the Ly$\alpha$ absorbers.  The newly gained
understanding of the nature of the Ly$\alpha$ forest and the increasing
realism of the simulations, together with new  semi-analytic methods
and novel ways of data analysis have brought quantitative cosmology
with the low column density Ly$\alpha$ forest within reach.

Aside from the cosmic microwave background, the intergalactic medium is
the only astrophysical environment for which observable properties  can (at
least in principle) be calculated from a simple set of cosmological
initial conditions. This is because at z$\sim
3$ the density fluctuations $\delta=\rho/\overline{\rho} - 1 $ on
spatial scales relevant for detectable Ly$\alpha$ lines (on the order of
10$^2$ kpc comoving) are not too far into the non-linear regime, so the
history of the gas causing most of the low column density Ly$\alpha$
forest has not yet been obliterated by virialization and dissipative
processes.  Overdensities between $\delta \stackrel{<}{\sim}0$ and
$\delta \sim 15$ roughly correspond to Ly$\alpha$ lines on the linear
part of the curve of growth ($12\simlt \log N \simlt 14$), where spectroscopic
measurements are most sensitive.  When observing structures still dominated
by gravity  the problem of ``bias'', one of the
main obstacles to doing cosmology with galaxies, largely can be avoided.

The link between the observable appearance of the Ly$\alpha$ forest
and the various cosmological input parameters can be described by the
Gunn-Peterson relation for the HI optical depth, generalized  to
include an inhomogenous density and velocity field.
As long as the gas is highly ionized and in photoionization equilibrium
(not necessarily thermal equilibrium), and the gas is unshocked, the
optical depth is given by
\begin{eqnarray}
\tau = \frac{0.091}{\Gamma_{-12}}\left(\frac{\Omega_b h_{50}^2}{0.05}\right)^{2}\left(\frac{5.2\ H(0)}{h_{50}H(z)}\right)T_{4}^{-0.7} \left(\frac{\rho}{\overline{\rho}}\right)^{\alpha} \left(\frac{1+z}{1+2}\right)^6 \left(1+\frac{dv_{pec}}{H(z)dr}\right)^{-1} \label{taueq}
\end{eqnarray}
This equation relates the optical depth for Ly$\alpha$ absorption to
the mean baryonic density (in gas) in units of the critical density,
$\Omega_b$, the Hubble constant at redshift $z$, $H(z)$, the average gas
temperature $T$, the proper baryon density $\rho$, the photoionization
rate $\Gamma_{-12}$ in units of 10$^{-12} s^{-1}$, and the gradient of
the local peculiar velocity $dv_{pec}/dr$ along the LOS.  A further
convolution with a Voigt profile is necessary to include the proper
thermal velocity broadening.  To turn this relation into a complete
description of the observed Ly$\alpha$ forest, cosmology has to predict
the cosmic density and velocity fields, the fraction of the closure
density in the form of gas, the equation of state, $T=T(\rho)$, and
the ionizing radiation field.
The exponent $\alpha$ ($\alpha$=2 for an isothermal gas)
takes account of the fact that in denser regions of the universe the
gas is typically warmer because it is more effectively heated by
photoionization, but $\alpha$ also depends on the reionization history
of the gas and the amount of adiabatic expansion/compression.  Hui \&
Gnedin (1997), and  Croft et al (1997a) find values of $\alpha\approx
1.6-1.8$.

Cosmological parameters can now be ``measured'' by
iterating simulations with different input parameters until the
simulated statistics of $\tau(z)$ (the mean absorption, correlation
function etc) match the observed ones.  Perhaps the simplest
cosmological parameter combination to be obtained is related to the
baryon density, $\Omega_b$. Given eqn.  (\ref{taueq}) the optical depth
$\tau$ scales approximately with $(\Omega h_{50}^2)^{\alpha
}/\Gamma_{-12}$.  Using an independent estimate of the photoionization
rate $\Gamma_{-12}$, e.g., from the integrated UV radiation of QSOs,
one can determine $\Omega_b$.

The simulations show that a rather high $\Omega_b h^2$ is required to
reproduce the amount of observed absorption,  even for a conservatively
low estimate of $\Gamma$ (Hernquist et al 1996, Miralda-Escud\'e et al
1996).  If the ionizing background is given by the known QSOs alone and
a Haardt \& Madau (1996) spectrum (scaled to $J_{-21} \approx 0.23$, at
$z$=2) is adopted, $\Omega_b h^2 >$ 0.017 (Rauch et al 1997). 
Although the observations of $\tau_{\mathrm eff}$ on which the result
is based, are still somewhat uncertain, relatively 
large $\Omega_b$ values appear to be an inherent feature of hierarchical
structure formation and cannot be avoided unless most of the low column
density absorption dominating the Ly$\alpha$ forest has a different
physical origin (Weinberg et al 1997).

\medskip

{\noindent\it \small COSMOLOGICAL PARAMETERS WITHOUT HYDRODYNAMICS\ \ }
Rerunning the simulations with different input parameters and/or
evolving them over a long time span is expensive, as is the analysis of
large observed and simulated datasets.  A number of new
semi-analytical techniques have been developed to avoid these
difficulties. Although not ``hydrodynamically correct'' they can give
new insights into important aspects of the sometimes obscure dependence
of the observational properties on the underlying physical
environment.  Such techniques compensate for the absent hydrodynamics
by various analytical recipes:  Petitjean et al (1995), and Croft et
al (1997b) used dark matter simulations and assume that baryons (furnished 
with a suitable analytical thermal history) trace the dark matter
directly; Bi et al (1992), and Hui et al (1997)
applied power spectra with various forms of cutoff to mimic the smoothing
introduced by the Jeans length; Gnedin \& Hui (1997), using a dark
matter code, simulate the effects of the gas pressure by modifying the
gravitational potential.

\medskip

{\noindent\it \small THE LYMAN ALPHA FOREST AS A RECORD OF PRIMORDIAL
FLUCTUATIONS\ \ } Semianalytical work by Gnedin \& Hui (1996) and
Hui et al (1997) has elucidated the relation between column
density peaks (``absorption lines'') and the statistics of density
peaks, and has given analytical expressions for the dependence of the
shape of the column density distribution function $f(N)$ on
cosmological parameters.  The slope of $f(N)$ was found
largely to be determined by the normalization and the slope of the initial
mass power spectrum, with changes in the equation of state $T=T(\rho)$
having an additional, but smaller impact. The overall normalization of
$f(N)$ is given by $(\Omega_b h^2)^2/\Gamma$; a change in this quantity
shifts $f(N)$ horizontally to larger $N$.
  
Croft et al (1997b) have suggested a technique for recovering the
initial power spectrum of density fluctuations directly from the
fluctuations of the optical depth. The distribution function of pixel
fluxes in a Ly$\alpha$ forest spectrum on larger scales is assumed to
have originated via gravitational collapse from an initially Gaussian
probability distribution of overdensities $\delta$.  Then the flux
probability function can be mapped monotonically according to the rank of
the flux values back onto a Gaussian probability function for the initial
$\delta$.  The density power spectrum (width of the Gaussian) $P(k)$ is
then known up to a normalization, which can be derived from iterative
cosmological simulations with the same $P(k)$ but different
normalization until the right flux power spectrum is obtained from
simulated spectra.  The flux power spectrum is unique up to a
normalization which can be fixed by observing the mean absorption
$\overline{D}$.

\medskip

{\noindent\it\small POTENTIAL PROBLEMS - UNSOLVED QUESTIONS\ \ } The
new paradigm for the Ly$\alpha$ forest has considerable explanatory
power, but that does not mean that it is correct. The interpretation of the absorption
systems, and the cosmological measurements planned or already performed
to date depend on gravitational collapse as the dominant source of
structure in the intergalactic medium. Even if the hierarchical models
are basically correct, it is conceivable that
local physical effects may upset some of the
cosmological conclusions.  A fluctuating radiation field may be a
source of non-gravitational structure in the forest, as may be stellar
feedback.  How much of the absorption is caused by gas blown out by
supernova explosions or stellar winds, and how robust are the
cosmological conclusions in that case ?  Metal enrichment has been
found to be common for absorption systems with HI column densities as
low as $\log N(HI)\sim 14$ (Tytler et al 1995; Cowie et al 1995).  If
this is not due to a very early phase of metal enrichment we have to
worry that some process other than gravitational collapse may have
formed the metal enriched Ly$\alpha$ clouds.  The origin of the ionizing radiation
and the spectral slope are another source of uncertainty.  When and how
do reionization and reheating happen, and where do the photons come
from ? Is collisional reheating important, and how much do stars
contribute to the UV flux, as a function of time ?
On the technical side: Do hydrodynamic codes already converge, or how
much do the inferred cosmological parameters ($\Omega_b$, the amount of
small scale structure present) depend on the resolution and size of the
simulations, and the numerical technique ? 

Finally, the cosmological picture itself could be wrong, 
and the interpretation of the forest as absorption mostly by the
intergalactic medium (as opposed to distinct galaxies) may be doubtful.
Galaxy halos or disks could be big/numerous enough to produce the
low column density Ly$\alpha$ forest as well. For low redshift absorption lines, this
last possibility has received much attention, and we will briefly consider
this question next.

\section{\large THE LYMAN ALPHA FOREST AND GALAXIES}

Is the Ly$\alpha$ forest absorption caused by galaxies ?  To match the
large rate of incidence of typical Ly$\alpha$ absorbers, ``normal''
(i.e., known types of) galaxies must possess very large absorption
cross-sections (Burbidge et al 1977).  Alternatively, since the rate of
incidence is constraining only the product of number density and
geometric cross-section of the absorbers, there could be a population
of unknown, more numerous objects, subtending a smaller cross-section.
Two key observations have fuelled the interest in the nature of the
absorber-galaxy connection. One was the detection of apparently
ordinary galaxies at the same (intermediate) redshifts as high column
density, MgII metal absorbers (Bergeron 1986).  Subsequent work
(Bergeron \& Boiss\'e 1991; Steidel 1995) established an
incontrovertible connection between gaseous galactic halos and high
column density, metal absorption systems (which we will not discuss here,
as it is most relevant for optically thick Lyman limit systems).  The
other was the widely unanticipated detection of a remnant population of Ly$\alpha$ absorbers in the
local universe with HST (Morris 1991; Bahcall et al 1991,1993), where the
properties of an galaxy can be studied in much greater detail than at
high redshift.

\subsection{The Low Redshift Lyman Alpha Forest}

Soon after the discovery of the low z absorbers, galaxy surveys in the
fields of the QSOs observed by HST, especially near 3C273, were undertaken to
investigate a possible link between absorbers and galaxies. Salzer (1992),
Morris et al (1993) and Salpeter \& Hoffman (1995)  found redshift
coincidences with galaxies for some of the lines, hundreds of kpc up
to Mpcs away from the absorbers, but there was no unambiguous
correspondence between absorbers and individual galaxies.  Searches for
any low surface brightness objects closer to the LOS to 3C273 which
could have escaped detection, have been performed in HI radio emission
(van Gorkom et al 1993), H$\alpha$ emission (Morris et al 1993, Vogel et
al 1995), and in deep broad band optical images (Rauch et al (1996), but were all 
equally unsuccessful. Morris et al (1993)
concluded that there is some correlation between absorbers and
galaxies, but the galaxy-absorber cross correlation function is weaker
than the galaxy-galaxy correlation.  By trying to model the correlation
results as a mixture of randomly distributed objects and galactic
halos, Mo \& Morris (1994) deduced that galaxy halos constitute only about
25\% of the local absorber population.  At first glance this is at variance with the
result of Lanzetta et al (1995), who, based on their large galaxy
redshift survey, concluded that most absorption systems are associated
with galactic envelopes of typical radius 160 $h^{-1}$ kpc (see also
Chen et al 1997).  Moreover, a significant anticorrelation between the
equivalent width and the impact parameter of the LOS from the center of
the galaxies was measured.  Le Brun et al (1996)'s
survey showed a weaker anticorrelation, and Bowen et al
(1996) saw none at all, although they both confirm Lanzetta et al's results in
that there is a region with a  high covering factor for absorption around
each galaxy, within impact parameters $<200-300 h^{-1}$ kpc, dropping
rapidly at larger separations.  It appears that the conclusions of
Morris et al (1991) and Lanzetta et al (1995) can be reconciled if the
Lanzetta et al's main result, i.e., the existing of well defined galactic
envelopes, is valid only for the relatively strong absorption lines
used in their comparison.  The weaker lines, from which  Morris et al
(1991) derived a weak galaxy-absorber correlation, could still be
caused by a truly intergalactic medium.  Absorption systems do exist in
voids known from galaxy surveys (Stocke et al 1995; Shull et al 1996), but
there appears to be a general trend for the absorbers to trace the same
large scale structure as galaxies (Stocke et al.  1995; Hoffman et al 1995).  In any case, the very large structures causing the
stronger common absorption systems in QSO pairs (Bechtold 1994; Dinshaw
et al 1994, 1995) cannot be explained by single objects like giant disks or
halos:  in those cases where several galaxies have been found at the same
redshift as the absorber the typical transverse separation between the
galaxies is smaller than the transverse absorber size (Rauch et al 1996). This is reminiscent of Oort's (1981) ``uncondensed gas
in superclusters''.  There is amazing redshift agreement (velocity
differences $0<\Delta v< 20$ kms$^{-1}$) between the HI velocity centroids of 
galaxies and
absorption systems with impact parameters as large as 300 $h^{-1}$ kpc
(van Gorkom et al 1996).

\medskip

{\it\noindent\small MODELLING LOW REDSHIFT LYMAN ALPHA ABSORBERS\ \ }
Attempts at modelling have mainly been concerned with the large
cross-section required if the low redshift Ly$\alpha$ clouds are parts
of the known galaxy population.  Maloney (1992) has suggested that the
absorption arises in the distant, ionized outer parts of the known
population of disk and irregular galaxies.  Clouds that are pressure confined by
a hotter gas in a extended galactic halo were discussed by Mo (1994).
Salpeter (1993) has invoked a new type of extended low redshift disk
galaxies which, in a short burst of star formation blow out their
denser centers and disappear from view, while their gas cross-section
remains (the so-called ``Cheshire Cat model'').  The large sizes and
low column densities postulated require that these objects must have
formed late (Salpeter \& Hoffman 1995).  Again, observationally the
large coherence lengths of the absorbers on the sky seem to correspond
to groups of galaxies. To get the large covering factor right these
objects must be so close together as to run into each other like
circular saws at only slightly higher redshift, which puts us basically back at a continuous intergalactic medium.  A number of other
sources of low z Ly$\alpha$ absorption has been considered, among them
tidal tails (Morris \& Van den Berg 1994), and galactic winds (Wang
1995). Galaxy clusters have been seen to cause at least some absorbers
(Lanzetta et al 1996), and the strong clustering found among  low z
absorption lines (Boksenberg 1995; Bahcall et al 1996; Ulmer 1996)
hints at the increasing importance of such structures with decreasing
redshift.

Clearly, one has to be open-minded regarding all these
possibilities, many of which plausibly may contribute a significant
fraction to low z Ly$\alpha$ absorption. How much exactly may be hard
to quantify (Sarajedini et al 1996). Again, believers in
hierarchical structure formation can expect some consolation from
simulations.  Petitjean et al  (1995) predicted a bi-modal distribution of absorbers, with large galaxy halos
with typical radii on the order of 0.5 $h^{-1}$ Mpc on one hand, and more frequent,
lower column density intergalactic absorption occuring in filaments up to
several Mpc away from the nearest galaxy, on the other.  Miraculously, this is
consistent with all the observational evidence we have.

\subsection{Metal Enrichment In The High Redshift Ly$\alpha$ Forest}

True to our self-imposed restraints we will deal here only with the low
column density forest. For the reasons already mentioned (large
transverse sizes, small velocity differences over hundreds of kpc,
probable flattening, absence of voids, weak LOS correlation, low
Doppler parameters, and the low column densities themselves) the
typical Ly$\alpha$ system at z$\sim 3$ is quite unlikely to be
physically associated with a galaxy, other than being part of the
gaseous matrix from which galaxies form. One puzzling result
remains, however: the finding of widespread metal enrichment 
in the Ly$\alpha$ forest.  

Individual high column density systems with very low metallicities (of
order $10^{-3}\odot$) exist (eg. Chaffee et 1985), but there is no
absorption system known with column density above $\log N(HI) > 16$ and
a primordial composition. The possibility that there may be a
transition at a certain column density from a metal enriched gas to a
primordial gas has led to intensive searches for weak metal lines in
low column density  absorbers.  Norris et al (1983), Williger et al
(1989), and Lu (1991) used a ``shift and stack'' method to search for
various metals in the low column density forest.  To maximize the
signal-to-noise ratio the spectra are shifted to the rest frame
indicated by each Ly$\alpha$, and added, and the expected positions of
metal lines are searched for a signal. A tentative detection of CIV,
corresponding to a carbon depletion relative to solar of [C/H]$\approx -3.1$ was made by Lu
et al 1991.  Meyer \& York (1987) pointed out, that data with
an increasing S/N ratio show increasing numbers of individual weak CIV
systems.  The subject attracted renewed interest when Keck spectra
showed that most Ly$\alpha$ systems with HI column densities $10^{15}$
and roughly half of all Ly$\alpha$ systems with column densities
$>3\times 10^{14}$ cm$^{-2}$ have associated CIV lines, corresponding
to a typical metallicity of $Z$$\sim 10^{-2} Z\odot$ (Cowie et al 1995; Tytler
et al 1995; Songaila \& Cowie 1996).  Unfortunately, the detection
threshold for CIV is close enough to make it hard to determine whether
the decreasing rate of detections is a genuine turnover to primordial
composition below a few $\times 10^{14}$, or just a selection effect.
It is interesting that in the CDM models a column density contour of
$10^{14}$ at redshift 3 delineates the transition between a continuous
filamentary structure in the universe, with typical widths of less than
100 kpc, and the voids.  Galaxies would have to spill metals only
within the filaments to create the widespread metal-enrichment
observed, and they could still have left most of the volume of the universe
pristine. Thus a drop in metallicity at a few $\times 10^{14}$ would
not come unexpected.  A relatively uniform metal abundance across the
whole column density range could be an indication of a earlier phase of
nucleosynthesis.

\section{\large THE HELIUM LYMAN ALPHA FOREST}

Observations of absorption by the He I and HeII Lyman series provide
another, independent source of information on the state of the
intergalactic medium and the UV background radiation (Sargent et al
1980). Comparing the Doppler parameters of the HeII 304 \AA\ and HI
1215 \AA\ lines it is in principle possible to use the difference in atomic masses
to measure the contributions from bulk and thermal motions to the line
broadening separately, so theories of the kinematics of the cosmic gas
can be tested.  The far-UV HeII Lyman edge at 228 \AA\ probes the
intensity of the UV background at much shorter wavelengths than the HI
edge. As the photoionization rates are dominated by the intensity of
the ionizing radiation near the respective ionization edges,
measurements of the HeII and HI column densities can then in principle
fix the spectral shape of the UV background in the vicinity of two
points, 228 \AA\ and 912 \AA .  This comparison can be done as a
function of wavelength (or redshift) so it is possible to measure the
spatial spectral fluctuations of the UV background caused by local UV
sources and by fluctuations in the intergalactic absorption, and, perhaps at
higher redshifts, to study the progress of reionization (e.g., Shapiro \& Giroux 1987;
Donahue \& Shull 1987; Miralda-Escud\'e \& Rees 1993; Shapiro et al 1994; Madau \&
Meiksin 1994; Giroux et al 1995).

\medskip 
In a photoionized optically thin gas HeIII is the dominant 
ionization state; the remaining He is expected to be mostly in the form
of singly ionized HeII.  For realistic spectral intensity distributions
(except for the very soft UV background caused by decaying neutrinos, Sciama
1990) HeI is undetectable in the optically thin clouds of the
Ly$\alpha$ forest proper, although it should be present in Lyman limit
systems (Miralda-Escud\'e \& Ostriker 1992). Indeed, HeI yielded the first
detection of helium at high redshift, when absorption by its 504 \AA\
line was found in HST FOS data of several optically thick ($z\sim2$) metal systems towards
the bright QSO HS1700+64 (Reimers et al 1992, Reimers \& Vogel 1993).
The ionization state of HeI is less easy to interpret because of the
possible presence of internal sources of radiation, and unknown shielding
effects.
\smallskip

\subsection*{\it The HeII Ly$\alpha$ Forest} 

Because of its relative strength, HeII 304 \AA\ is likely to be
a better tracer of the low density baryon distribution than even
HI Ly$\alpha$.

The principal observable of the HeII Ly$\alpha$ forest is the ratio of the Gunn-Peterson optical depths of HeII to HI:
\begin{eqnarray}
\frac{\tau_{\rm HeII}}{\tau_{\rm HI}}=\frac{1}{4} \frac{N_{\rm HeII}}{N_{\rm HI}} = \frac{1}{4} \eta=0.43 \frac{J_{\rm HI}}{J_{\rm HeII}},
\label{taugphe}\end{eqnarray}
where $N_{\rm HeII}$ and $N_{\rm HI}$ are the column densities of HeII
and HI, respectively (Miralda-Escud\'e 1993). The last equation is valid
for an optically thin gas, where the ionization equilibrium is governed
by photoionisation and both H and He are highly ionized. $J_{\rm
HeII}$ and $J_{\rm HI}$ are the intensities of the ionizing radiation
field weighted with the frequency dependence of the photoionization
cross-sections. Thus the relative strength of the GP troughs are only
dependent on the ratios of the ionizing fluxes, a situation which can
be cast in terms of the column density ratio $\eta$ or the ratio of
the intensities at the absorption edges, the softness parameter $S_L$
(Madau \& Meiksin 1994),
\begin{eqnarray}
\eta =\frac{N_{\rm HeII}}{N_{\rm HI}}\approx 1.8 \frac{J_{\rm 912}}{J_{\rm 228}}= 1.8 S_L.
\end{eqnarray}
The HeII/HI ratio $\eta$ can range from values of a few, to a thousand,
depending on the spectral slope of the ionizing radiation. 
The shape of the ionizing spectrum depends on the relative mix between ``hard''
(AGNs) and ``soft'' (stellar) sources, and on the details of radiative
transfer by the intergalactic medium (Bechtold et al 1987; 
Shapiro \& Giroux 1987; Miralda-Escud\'e \& Ostriker 1990;
Meiksin \& Madau 1993; Giroux \& Shapiro 1996, Haardt \& Madau 1996).

Thus a HeII 304 \AA\ Gunn-Peterson trough should
appear more prominent than the corresponding HI trough (Miralda-Escud\'e
1993), allowing for a more sensitive measurement of the distribution of
gas in low density regions. 
For the actual HeII forest of discrete absorption lines a relation similar to equation (\ref{taugphe})
holds, where $\tau$ is now replaced by $\tau_{\mathrm eff}$ as defined
earlier (Miralda-Escud\'e 1993):
\begin{eqnarray}
\frac{\tau_{\rm eff\ HeII}}{\tau_{\rm eff\ HI}}=\left(0.43 \frac{J_{\rm HI}}{J_{\rm HeII}}\right)^{\beta-1}\left(\frac{b_{\rm HeII}}{b_{\rm HI}}\right)^{2-\beta}.
\label{taulhe}\end{eqnarray} 
This relation gives the optical depths for a Ly$\alpha$ forest of individual lines above a certain optical depth threshold, assuming that the column density distribution of HI Ly$\alpha$ lines is a power law with index $\beta$. The Doppler parameters are $b_{\rm HeII}$ and $b_{\rm HI}$.
Clearly, together with the strong HeII GP effect  there should be a HeII
forest stronger than the corresponding HI Ly$\alpha$ forest by a similar
factor. 

\medskip

{\it OBSERVATIONS \ } Only a small fraction of all QSOs are suitable for
a search for HeII absorption. The short wavelength of 304 \AA\ requires
an object to be redshifted to at least $z>2-3$ for  the HeII line to
enter the far UV bands accessible to the Hubble Space Telescope (HST),
or the Hopkins Ultraviolet Telescope (HUT). The QSO has to be bright
enough for a spectrum to be taken, and most importantly, there must be
flux down to the wavelength range of interest. The average blanketing of
the spectrum by Ly$\alpha$ lines (the "Lyman valley", M\o ller \& Jakobsen
1990) and especially the total blackout imposed by individual intervening
Lyman limit systems below 912 \AA\ in their rest frame renders the large
majority of QSOs useless for a HeII search (Picard \& Jakobsen 1993), and
surveys of known QSOs for residual UV flux (e.g. Jakobsen et al 1993)
have to be mounted to select suitable candidates.

To date there are five detections of the HeII forest.  A HeII absorption
"break" blueward of the HeII emission line was first seen with the
HST FOC far UV prism by Jakobsen et al, 1994 in the LOS to Q0302-003
($z_{em}$=3.29), leading to an estimate for the mean optical depth,
$\tau_{\rm HeII} = 3.2^{+\infty}_{-1.1}$. The object was reobserved at higher resolution
with the GHRS instrument by Hogan et al (1997) ($\tau_{\rm
HeII} \approx 2$, beyond the proximity effect region).  Tytler et al. (1995)
obtained $\tau_{\rm HeII}=1.0\pm 0.2$, later corrected to $\tau_{\rm HeII}
> 1.5$ (Tytler \& Jakobsen, unpublished) in the LOS to Q1937-69 ($z_{em}$=3.18).
Davidsen et al (1996), observed the object HS1700+6416 with
HUT to obtain $\tau_{\rm HeII} = 1.0 \pm 0.2$, at $<z>$=2.4.  Reimers et
al (1997), in the LOS to HE2347-4342 ($z_{em}$=2.89) find the HeII
absorption to consist of patches with a high continuous GP component,
$\tau_{\rm HeII} = 4.8^{+\infty}_{-2}$ in addition to the contribution
expected from the discrete lines, which alternate with regions with less GP
absorption  $\tau_{\rm HeII} \approx 3$.

\medskip

{\it INTERPRETATION\ } Given the large uncertainties, all $\tau$
measurements to date seem to be consistent with each other, if the expected increase in the optical depth with redshift is taken into account.
Constraining the strength and shape of the ionizing radiation from the
absorbed flux requires a knowledge of the clumpiness of the gas,
because of the exponential dependence of the absorbed flux on the
optical depth.  According to equations (\ref{taugphe}) and
(\ref{taulhe}) the relative strengths of the absorption by HI and HeII
depend on the amount of bulk motion relative to thermal motion, and on
the column density distribution function (slope and possible cutoff at
low column density), and the relative contribution  from a diffuse
absorption trough.  Arguments have been put forward both against
(Songaila et al 1995) and in favor (Hogan et al 1997; Reimers et al
1997; Zheng et al 1998) of the existence of such a trough in addition
to the line absorption expected from translating the known HI
Ly$\alpha$ forest into a HeII forest. If the proponents of additional
trough absorption are correct this may also imply that the HI column
density distribution function has been over-corrected for confusion,
and does not extend to as low a column density as previously assumed.
Superficially this argument sounds like the return of the
lines-versus-trough debate familiar from the HI Gunn-Peterson effect,
but there is new information to be gained by studying the detailed
structure of the low density HeII absorption. The larger optical depth
of HeII highlights very low density structure in voids, which may be
too weak to be usefully constrained with optical HI Ly$\alpha$ forest
spectra.  Eventually such observations will constrain the spatial
fluctuations of the ionizing radiation field and the density field in a
large fraction of the volume of the the universe.  The question  of
whether the spatial variations of the diffuse absorption can be (or
should be) parametrized as ``lines'' may have to await the arrival of
better data. In any case, the finding of a substantial amount of HeII
absorption from voids is an important consistency check for
hierarchical structure formation models (Zhang et al 1995, 1997; Croft et al 1997a).

At the time of this writing the spectral shape of the radiation field
is still not well constrained (Sethi \& Nath 1997; Reimers et al 1997;
Zheng et al 1998). An interesting twist has been added by
the detection of patchy HeII absorption, which is inconsistent with a
uniform radiation field. Reimers et al (1997) invoke
incomplete reionization of HeII as a possible explanation, an effect
predicted to produce saturated absorption troughs (Meiksin \&
Madau 1993).  However, it is hard to tell how strongly saturated the troughs
really are. The observations may still be consistent with a fully
re-ionized HeII, if the troughs are caused by local fluctuations in the HeII ionizing background (Miralda-Escud\'e 1997).

\section{\large PROSPECTS}

Although most of the observational basis and many theoretical aspects
of Ly$\alpha$ forest absorbers had already been established over the
past two decades, high signal-to-noise spectroscopy of QSOs with large
telescopes, the extension of the wavelength range into the far UV with
satellites, and the success of hydrodynamic cosmological simulations
have begun to turn the study of the Ly$\alpha$ forest from  a somewhat
esoteric appendix of cosmology into a cosmologically useful tool. If
our current understanding is correct, the high redshift Ly$\alpha$
forest absorption is the observational signature of most of the baryons
throughout most of the history of the universe. Perhaps most
importantly, we are looking at the typical fate of matter, without any
reference to luminous objects.

The new general picture of the forest that has emerged is still in need
of more secure observational conformation, before we can trust any
quantitative cosmological conclusions.  The simulations agree with some
aspects of the data, but are they unique, and consistent with each
other ?  More detailed comparisons with the data are called for, partly
with new techniques of data analysis tailored to maximize the
discriminative power.  Semi-analytical work still has an important
function in guiding our understanding of the actual physics, and in
exploring parameter space quickly.  The observers need to explore the
limitations of the observational techniques, and systematic effects in
the data, formerly gracefully hidden by a veil of noise, but now
exposed by large mirrors to the strict eye of numerical theory.

We can expect new observational facilities to enlarge the scope of
absorption line studies considerably. The projected Cosmic Origins
Spectrograph (COS) will increase the spectroscopic efficiency of the
HST, benefitting almost all of the observational areas mentioned here.
The Sloan Digital Sky Survey (SDSS) should produce large numbers of
QSOs useful for spectroscopic follow-up, to study (for example) the
large scale structure from intergalactic absorption in three
dimensions. Even low resolution QSO spectra will be useful for
cosmological purposes.  And the use of large optical telescopes is
inevitable for studies of gravitationally lensed QSOs, the
absorber-galaxy connection at any redshift, and of course for most
projects involving narrow, metal absorption lines, a subject of
increasing relevance for our understanding of the process of galaxy
formation.

Unfortunately, other applications of absorption lines like the
remarkably militant debate about the deuterium to hydrogen ratio,
the whole topic of intervening metal absorption systems and its relevance for
galaxy formation, and the process of reionization could not be treated
here.  Perhaps we have ignored them with some justification; these subjects are 
currently in such a state of flux that any review could be out of date before
going to press.

\subsection*{ACKNOWLEDGMENTS}

I am  grateful to Bob Carswell, Martin Haehnelt,  Jordi
Miralda-Escud\'e, and Wal Sargent for reading earlier drafts, to Naomi
Lubick and Allan Sandage for their thorough editorship, and to NASA for
support through grant HF-01075.01-94A from the Space Telescope Science
Institute, which is operated by the Association of Universities for
Research in Astronomy, Inc., under NASA contract NAS5-26555.

\pagebreak

\section{\it Literature Cited}

\noindent Acharya M, Khare P. 1993. {\it JAp\& A} 14:97

\noindent Arons J, McCray R. 1969. {\it Ap Letters} 5:123

\noindent Arons J. 1972. {\it ApJ} 172:553


\noindent Atwood B, Baldwin RA, Carswell RF. 1985. {\it ApJ} 292:58

\noindent Babul A. 1990. {\it ApJ} 349:429

\noindent Babul A. 1991. {\it MNRAS} 248:177

\noindent Babul A, Rees MJ. 1992. {\it MNRAS} 255:346

\noindent Bahcall JN, Salpeter EE. 1965. {\it ApJ} 142:1677

\noindent Bahcall JN, Peebles PJE 1969. {\it ApJ} 156:L7

\noindent Bahcall JN, Spitzer L. 1969. {\it ApJ} 156:L63

\noindent Bahcall JN, Jannuzi BT, Schneider DP, Hartig GF, Bohlin R, Junkkarinen V.
1991. {\it ApJLetters} 377:5

\noindent Bahcall JN, Bergeron J, Boksenberg A, Hartig GF, Jannuzi BT. 1996. {\it ApJ} 457:19

\noindent Bajtlik S, Duncan RC, Ostriker JP. 1988. {\it ApJ} 327:570

\noindent Baldwin JA, Burbidge EM, Burbidge GR, Hazard C, Robinson LB, et al. 1974. {\it ApJ} 193:513

\noindent Barcons X, Fabian AC. 1987. {\it MNRAS} 224:675

\noindent Barcons X, Webb JK. 1990. {\it MNRAS} 244:30p 

\noindent Barcons X, Webb JK. 1991. {\it MNRAS} 253:207

\noindent Barcons X, Fabian AC, Rees MJ. 1991 {\it Nature} 350:685

\noindent Baron E, Carswell RF, Hogan CJ, Weymann RJ. 1989. {\it ApJ} 337:609

\noindent Black J. 1981. {\it MNRAS} 197:553

\noindent Bechtold J, Green RF, Weymann RJ, Schmidt M, Estabrook F, et al. 1984.{\it ApJ}
281:76

\noindent Bechtold J, Weymann RJ, Zuo L, Malkan MA. 1987. {\it ApJ} 315:118

\noindent Bechtold J. 1987. in:{\it High Redshift and Primeval Galaxies}. eds.
Bergeron J, Kunth D, Rocca-Volmerange B, Tran Thanh Van J.

\noindent Bechtold J. 1994. {\it ApJS} 91:1

\noindent Bechtold J, Crotts APS, Duncan RC, Fang Y. 1994 {\it ApJ} 437:83

\noindent Bechtold J. 1995. in {\it QSO Absorption Lines}. p 299, ed. Meylan G. Berlin: Springer

\noindent Bergeron J. 1986. {\it A\&A} 155:L8

\noindent Bergeron J, Boiss\'e P. 1991. {\it A\&A} 243:344

\noindent Bi H, B\"orner G, Chu Y. 1989. {\it A\& A} 218:29

\noindent Bi H, B\"orner G, Chu Y. 1991. {\it A\& A} 247:276

\noindent Bi H, B\"orner G, Chu Y. 1992. {\it A\&A} 266:1

\noindent Bi H. 1993. {\it ApJ} 405:479

\noindent Bi H, Davidsen AF. 1997. {\it ApJ} 479:523

\noindent Boksenberg A. 1995. in {\it QSO Absorption Lines}. p 253; ed. Meylan G. Berlin: Springer

\noindent Bond JR, Szalay AS, Silk J. 1988. {\it ApJ} 324:627

\noindent Bond JR, Kofman L, Pogosyan D. 1996. {\it Nature} 380:603

\noindent Bond JR, Wadsley JW. 1997. in {\it Computational Astrophysics. Proc. 12th Kingston
Conference} Halifax Oct. 1996. p323. ed D. Clark, M. West. PASP conf. series.

\noindent  Bowen DV, Blades J, Pettini M. 1996. {\it ApJ} 464:141

\noindent Burbidge EM, Lynds CR, Burbidge GR. 1966. {\it ApJ} 144:447

\noindent Burbidge G, O'Dell SL, Roberts DH, Smith HE. 1977. {\it ApJ} 218:33

\noindent Carswell RF, Whelan JAJ, Smith MG, Boksenberg A, Tytler D. 1982. {\it MNRAS} 198:91

\noindent Carswell RF, Morton DC, Smith MG, Stockton AN, Turnshek DA, Weymann RJ. 1984. {\it ApJ} 278:486

\noindent Carswell RF, Rees, MJ. 1987. {\it MNRAS} 224:13

\noindent Carswell RF, Webb JK, Baldwin JA, Atwood B. 1987. {\it ApJ} 319:709

\noindent Carswell RF. 1988.  QSO Absorption Lines: Probing the Universe.
ed JC Blades, DA Turnshek, CA Norman.
{\it Proceedings of the QSO Absorption Line Meeting, Baltimore 1987}
Cambridge: Cambridge University Press 1988

\noindent Carswell RF, Lanzetta KM, Parnell HC, Webb JK. 1991. {\it ApJ} 371:36

\noindent Cen R, Ostriker JP,  1993. {\it ApJ} 417:404

\noindent Cen R, Miralda-Escud\'e J, Ostriker JP, Rauch M. 1994. {\it ApJLetters} 437:9

\noindent Cen R, Simcoe RA. 1997. {\it ApJ} 483:8

\noindent Chaffee FH, Weymann RJ, Strittmatter PA, Latham DW. 1983 {\it ApJ} 252:10

\noindent Chaffee FH, Foltz CB, Weymann RJ, R\"oser H-J, Latham DW. 1985. {\it ApJ} 292:362

\noindent Charlton JC, Salpeter EE, Hogan CJ. 1993. {\it ApJ} 402:493

\noindent Charlton JC, Salpeter EE, Linder SM. 1994. {\it ApJ} 430:29

\noindent Charlton JC, Anninos P, Zhang Y, Norman M. 1997. {\it ApJ} 485:26

\noindent Chen H-W, Lanzetta KM, Webb JK, Barcons X. 1998. {\it ApJ} in press

\noindent Chernomordik VV, \& Ozernoy LM. 1983. {\it Nature} 303:153

\noindent Chernomordik VV. 1988. {\it SvA} 32:6

\noindent Chernomordik VV, Ozernoy LM. 1993. {\it ApJ} 404:5

\noindent Chernomordik, VV. 1994. {\it ApJ} 440:431

\noindent Cooke AJ, Espey B, Carswell RF. 1997. {\it MNRAS} 284:552

\noindent Cowie LL, Songaila A, Kim T-S, Hu EM. 1995. {\it AJ} 109:1522

\noindent Cristiani S, D'Odorico S, Fontana A, Giallongo E,
Savaglio S. 1995 {\it MNRAS} 273:1016

\noindent Cristiani S, D'Odorico S, D'Odorico V, Fontana A, Giallongo E,
Savaglio S. 1997 {\it MNRAS} 285:209

\noindent Croft RA, Weinberg DH, Katz N, Hernquist L. 1997. {\it ApJ} 488:532

\noindent Croft RA, Weinberg DH, Katz N, Hernquist L. 1998. {\it ApJ} 495:44

\noindent Crotts APS. 1987. {\it MNRAS} 228:41

\noindent Crotts APS. 1989. {\it ApJ} 336:550

\noindent Dav\`e R, Hernquist L, Weinberg DH, Katz N. 1997. {\it ApJ} 477:21

\noindent Davidsen AF, Kriss GA, Zheng W. 1996. {\it Nature} 380:47

\noindent Dinshaw N, Impey CD, Foltz CB, Weymann RJ, Chaffee FH. 1994. {\it ApJ} 437:87

\noindent Dinshaw N, Foltz CB, Impey CD, Weymann RJ, Morris SL. 1995. {\it Nature} 373:223

\noindent Dobrzycki A, Bechtold J. 1991. {\it ApJLetters} 377:69


\noindent Donahue M, Shull JM. 1987. {\it ApJ} 323:L13

\noindent Donahue M, Shull JM. 1991. {\it ApJ} 383:511

\noindent Doroshkevich AG, Shandarin SF. 1977. {\it MNRAS} 179:95

\noindent Doroshkevich AG, Muecket JP. 1985 {\it SvAL} 11:331

\noindent Duncan RC, Ostriker JP, Bajtlik S. 1989. {\it ApJ} 345:39

\noindent Duncan RC, Vishniac ET, Ostriker JP. 1991. {\it ApJ} 368:1

\noindent Espey BR. 1993. {\it ApJ} 411:59

\noindent Fang LZ. 1991 {\it A\&A} 244:1

\noindent Fang Y, Duncan RC, Crotts APS, Bechtold J. 1996 {\it ApJ} 462:77

\noindent Fardal MA, Shull JM. 1993. {\it ApJ} 415:524

\noindent Fernandez-Soto A, Barcons X, Carballo R, Webb JK. 1995. {\it MNRAS}
277:235

\noindent Fernandez-Soto A, Lanzetta KM, Barcons X, Carswell RF, Webb JK,
Yahil A. 1996. {\it ApJ} 460:85

\noindent Foltz CB, Weymann RJ, R\"oser HJ, Chaffee FH. 1984. {\it ApJ} 281:1

\noindent Francis PJ, Hewett PC. 1993. {\it AJ} 105:1633

\noindent Fransson C, Epstein R. 1982. {\it A\& A} 198:1127

\noindent Goldreich P, Sargent WLW. 1976. {\it CommAp} 6:133

\noindent Giallongo E. 1991. {\it MNRAS} 251:541

\noindent Giallongo E, Cristiani S. 1990. {\it MNRAS} 247:696

\noindent Giallongo E, Cristiani S, Tr\`evese D. 1992 {\it ApJ} 398:9

\noindent Giallongo E, Cristiani S, Fontana A, Tr\`evese D. 1993. {\it ApJ} 416:137

\noindent Giallongo E, D'Odorico S, Fontana A, McMahon R, Savaglio S. et al. 1994. {\it ApJLetters} 425:1

\noindent Giallongo E, Petitjean P. 1994. {\it ApJ} 426:61

\noindent Giallongo E, Cristiani S, D'Odorico S, Fontana A, Savaglio S.
1996. {\it ApJ} 466:46

\noindent Giroux ML, Fardal MA, Shull JM. 1995. {\it ApJ} 451:477

\noindent Giroux ML, Shapiro P. 1996. {\it ApJS} 102:191

\noindent Gnedin NY, Hui L. 1996. {\it ApJLetters} 472:73

\noindent Gnedin NY, Hui L. 1998. {\it MNRAS} in press.

\noindent Gunn JE, Peterson BA. 1965 {\it ApJ} 142:1633

\noindent Haardt F, Madau P. 1996. {\it ApJ} 461:20

\noindent Haehnelt M, Steinmetz M, Rauch M. 1996. {\it ApJ} 465:95

\noindent Haehnelt M, Rauch M, Steinmetz M. 1996. {\it MNRAS} 283:1055

\noindent Haehnelt M, Steinmetz M. 1998. {\it MNRAS}, in press.

\noindent Hernquist L, Katz N, Weinberg D, Miralda-Escud\'e. 1996. {\it ApJLetters} 457:51

\noindent Hoffman GL, Lewis BM, Salpeter EE. 1995. {\it ApJ} 441:28

\noindent Hogan CJ, Anderson SF, Rugers MH. 1997. {\it AJ} 113:1495

\noindent Hu EM, Kim T-S, Cowie LL, Songaila A, Rauch M. 1995. {\it AJ} 110:1526

\noindent Hui L, Gnedin NY, Zhang Y. 1997. {\it ApJ} 486:599

\noindent Hui L, Gnedin NY. 1997 {\it MNRAS}, 292:27

\noindent Hunstead RW, Pettini M, Blades JC, Murdoch HS, 1986. in:{\it Proceedings
of the 124th IAU Symposium}, Beijing. ed. A Hewitt, G Burbidge, Z-F Fang. Dordrecht:Reidel.

\noindent Hunstead RW, Murdoch HS, Pettini M, Blades JC. 1988. {\it ApJ} 329:527

\noindent Hunstead RW, Pettini M. 1991. in Shaver PA, Wampler EJ, Wolfe AM, eds.
{\it Proceedings of the ESO Mini-Workshop on QSO Absorption Lines}
ESO Scientific Report No. 9, 11. Garching:ESO

\noindent Ikeuchi S. 1986. {\it Ap\&SS} 118:509

\noindent Ikeuchi S, Ostriker JP. 1986. {\it ApJ} 301:522

\noindent Ikeuchi S, Murakami I, Rees MJ. 1988. {\it MNRAS} 236:21p

\noindent Impey CD, Petry CE, Malkan MA, Webb W. 1996. {\it ApJ} 463:473

\noindent Jakobsen P, Albrecht R, Barbieri C, Blades JC, Boksenberg A, et al. 1993. {\it ApJ} 417:528

\noindent Jakobsen P, Boksenberg A, Deharveng JM, Greenfield P, Jedrzejewski R,
Paresce F. 1994. {\it Nature} 370:35

\noindent Jenkins EB, Ostriker JP. 1991. {\it ApJ} 376:33

\noindent Katz N, Weinberg DH, Herquist L, Miralda-Escud\'e J. 1996. {\it ApJLetters} 457:57

\noindent Kim T-S, Hu EM, Cowie LL, Songaila A. 1997. {\it AJ} 114:1

\noindent Kinman TD. 1966. {\it ApJ} 144:1232

\noindent Kirkman D, Tytler D. 1997. {\it ApJ} 484:672

\noindent Kovner I, Rees MJ. 1989. {\it ApJ} 345:52

\noindent Kulkarni VP, Fall SM. 1993. {\it ApJ} 413:63

\noindent Kulkarni VP, Huang K, Green RF, Bechtold J, Welty DE, York DG. 1996.{\it MNRAS} 279:197

\noindent Lanzetta KM. 1988. {\it ApJ} 332:96

\noindent Lanzetta KM, Bowen DB, Tytler D, Webb JK. 1995. {\it ApJ} 442:538

\noindent Lanzetta KM, Webb JK, Barcons X. 1996. {\it ApJLetters} 456:17

\noindent LeBrun V, Bergeron J, Boiss\'e P. 1996. {\it A\& A} 306:691

\noindent Liu XD, Jones BJT. 1988. {\it MNRAS} 230:481

\noindent Liu XD, Jones BJT. 1990. {\it MNRAS} 242:678

\noindent Loeb A, Eisenstein DJ. 1995. {\it ApJ} 448:17

\noindent Lu L, Wolfe AM, Turnshek DA. 1991. {\it ApJ} 367:19

\noindent Lu L. 1991. {\it ApJ} 379:99

\noindent Lu L, Zuo L. 1994. {\it ApJ} 426:502

\noindent Lu L, Sargent WLW, Womble DS, Takada-Hidai M. 1996. {\it ApJ}
472:509

\noindent Lynds CR. 1970. IAU Symposium 44:127

\noindent Lynds CR, Stockton AN. 1966. {\it ApJ} 144:446

\noindent Madau P. Meiksin A. 1991. {\it ApJ} 374:6

\noindent Madau P, Meiksin A. 1994. {\it ApJ} 433:L53

\noindent Maloney P. 1992. {\it ApJ} 398:89

\noindent McGill C. 1990. {\it MNRAS} 242:544

\noindent Meiksin A, Madau P. 1993. {\it ApJ} 412:34

\noindent Meiksin A. 1994. {\it ApJ}431:109

\noindent Meiksin A., Bouchet FR. 1995. {\it ApJLett} 448:85

\noindent Melott A. 1980. {\it ApJ} 268: 630

\noindent Meyer DM, York DG. 1987. {\it ApJ} 315:5

\noindent Milgrom M. 1988. {\it A\&A} 202:9

\noindent Miralda-Escud\'e J, Ostriker JP. 1990 {\it ApJ} 350:1

\noindent Miralda-Escud\'e J, Ostriker JP. 1992. {\it ApJ} 392:15

\noindent Miralda-Escud\'e J, Rees MJ. 1993. {\it MNRAS} 260:617

\noindent Miralda-Escud\'e J. 1993. {\it MNRAS} 262:273

\noindent Miralda-Escud\'e J, Rees MJ. 1994.{\it MNRAS} 266:343

\noindent Miralda-Escud\'e J, Cen R, Ostriker JP, Rauch M. 1996. {\it ApJ} 471:582

\noindent Miralda-Escud\'e J. 1997. {\it ApJ} submitted.

\noindent Mo HJ, Xia XY, Deng ZG, B\"orner G, Fang L-Z. 1992 {\it A\& A} 256:L23

\noindent Mo HJ, Miralda-Escud\'e J, Rees MJ. 1993. {\it MNRAS} 264:705

\noindent Mo HJ. 1994. {\it MNRAS} 269:49

\noindent Mo HJ, Morris SL. 1994 {\it MNRAS} 269:52

\noindent Mo HJ, Miralda-Escud\'e J. 1996. {\it ApJ} 469:589

\noindent M\o ller P,  Jakobsen p. 1990. {\it A\& A} 228:299

\noindent M\o ller P, Kjaergaard P. 1991. {\it A\&A} 258:234

\noindent Morris SL, Weymann RJ, Savage BD, Gilliland RL. 1991. {\it ApJLetters}
377:21

\noindent Morris SL, Weymann RJ, Dressler A, McCarthy PJ, Smith BA, Terrile RJ,
Giovanelli R, Irwin M. 1993. {\it ApJ} 419:524

\noindent Morris SL, Van den Berg S. 1994. {\it ApJ} 427:696

\noindent Muecket JP, Mueller V. 1987. {\it Ap\&SS} 139:163

\noindent Muecket JP, Petitjean P, Kates RE, Riediger R. 1996. {\it A\&A} 308:17

\noindent Murakami I, Ikeuchi S. 1990. {\it PASJ} 42:L11

\noindent Murakami I, Ikeuchi S. 1993. {\it ApJ} 409:42

\noindent Murdoch HS, Hunstead RW, Pettini M, Blades JC. 1986. {\it ApJ} 309:19

\noindent Norris J, Hartwick FDA, Peterson BA. 1983. {\it MNRAS} 273:450

\noindent O'Brian PT, Gondhalekar PM, Wilson R. 1988. {\it MNRAS} 233:801

\noindent Oke JB, Korycansky DG, 1982. {\it ApJ} 255,11

\noindent Oort HJ. 1981. {\it A\&A} 94:359

\noindent Ostriker JP, Cowie LL. 1981. {\it ApJ} 243:127

\noindent Ostriker JP, Ikeuchi 1983. {\it ApJLetters} 268:63

\noindent Ostriker JP, Bajtlik S, Duncan RC. 1988. {\it ApJ} 327:350

\noindent Ozernoy LM, Chernomordik VV. 1976. {\it SvAL} 2:375

\noindent Pando J, Fang L-Z. 1996. {\it ApJ} 459:1

\noindent Paresce F, McKee CF, Bowyer S. 1980.{\it ApJ} 240:389

\noindent Parnell HC, Carswell RF. 1988. {\it MNRAS} 230:491

\noindent Peacock J. 1991. {\it Nature} 349:190

\noindent Peterson BA. 1978. in: Longair MS, Einasto J (eds.). {\it the Large
Scale Structure of the Universe} p. 389. D. Reidel. Dordrecht 1978.

\noindent Petitjean P, Bergeron J, Carswell RF, Puget JL. 1993. {\it MNRAS} 260:67

\noindent Petitjean P, Webb JK, Rauch M, Carswell RF, Lanzetta KM. 1993. {\it MNRAS} 262:499

\noindent Petitjean P, Rauch M, Carswell RF. 1994. {\it A\& A} 291:29p

\noindent Petitjean P, Muecket JP, Kates RE. 1995. {\it A\&A} 295:L12

\noindent Pettini M, Hunstead RW, Smith LJ, Mar DP. 1990. {\it MNRAS} 246:545

\noindent Picard A, Jakobsen P. 1993. {\it A\& A} 276:331

\noindent Pierre M, Shaver PA, Iovino A. 1988. {\it A\&A} 197:3

\noindent Press WH, Rybicki GB, Schneider DP. 1993. {\it ApJ} 414:64

\noindent Prochaska JX, Wolfe AM. 1997. {\it ApJ} 487:73

\noindent Rauch M, Carswell RF, Chaffee FH, Foltz CB, Webb JK, et al. 1992. {\it ApJ} 390:387

\noindent Rauch M, Carswell RF, Webb JK, Weymann RJ. 1993. {\it MNRAS} 260:589

\noindent Rauch M, Haehnelt MG. 1995 {\it MNRAS} 275:76

\noindent Rauch M, Weymann RJ, Morris SL. 1996. {\it ApJ} 458:518. 

\noindent Rauch M. 1996. in: {\it Cold Gas at High Redshift} eds. MN Bremer et al. p 137.
Dordrecht: Kluwer.

\noindent Rauch M, Miralda-Escud\'e J, Sargent WLW, Barlow TA, Weinberg DH, et al. 
1997, {\it ApJ} 489:7

\noindent Rees MJ, Setti G. 1970. {\it A\&A} 8:410

\noindent Rees MJ. 1986. {\it MNRAS} 218:25

\noindent Rees MJ. 1988.  QSO Absorption Lines: Probing the Universe.
ed. JC Blades, DA Turnshek, CA Norman.
{\it Proceedings of the QSO Absorption Line Meeting, Baltimore 1987}
Cambridge: Cambridge University Press 1988

\noindent Reimers D, Vogel S, Hagen HJ, Engels D, Groote D, et al. 1992. {\it Nature} 360:561

\noindent Reimers D, Vogel S. 1993. {\it A\& A} 276:L13

\noindent Reimers D, K\"ohler S, Wisotzki L, Groote D, Rodriguez-Pascal P, Wamsteker W. 1997. {\it A\& A} 327:890

\noindent Reisenegger A, Miralda-Escud\'e J. 1995. {\it ApJ} 449:476

\noindent Riediger R, Petitjean P, Muecket JP. 1998. {\it A\&A} 329:30.

\noindent Salpeter EE. 1993. {\it AJ} 106:1265

\noindent Salpeter EE, Hoffman GL. 1995. {\it ApJ} 441:51

\noindent Salzer JJ. 1992. {\it AJ} 103:385

\noindent Sarajedini V, Green RF, Jannuzi BT. 1996. {\it ApJ} 457:542

\noindent Sargent WLW, Young PJ, Boksenberg A, Carswell RF, Whelan JAJ. 1979 {\it ApJ} 230: 49

\noindent Sargent WLW, Young PJ, Boksenberg A, Tytler D. 1980 {\it ApJS} 42:41

\noindent Sargent WLW, Young PJ, Schneider DP. 1982. {\it ApJ} 256:374

\noindent Sargent WLW, Steidel CC, Boksenberg A. 1989. {\it ApJ} 69:703

\noindent Savaglio S, Cristiani S, D`Odorico S, Fontana A,
Giallongo E, Molaro P. 1997. {\it A\& A} 318:347

\noindent Scheuer PAG. 1965. {\it Nature} 207:963

\noindent Schneider D, Schmidt M, Gunn JE. 1991. {\it AJ} 102:837

\noindent Schwarz J, Ostriker JP, Yahil A. 1975 {\it ApJ} 202:1

\noindent Sciama D, 1990. {\it ApJ} 364:549

\noindent Sethi SK, Nath BB. 1997. {\it MNRAS} 289:634

\noindent Shapiro PR, Giroux ML. 1987. {\it ApJ} 321:L107

\noindent Shapiro PR, Giroux ML, Babul A. 1994. {\it ApJ} 427:25

\noindent Shaver PA, Robertson JG. 1983. {\it ApJ} 268:57

\noindent Shklovski IS. 1965 {\it SovA} 8:638

\noindent Shull JM, Stocke JT, Penton S. 1996. {\it AJ} 111:72

\noindent Smette A, Surdej J, Shaver PA, Foltz CB, Chaffee FH, Weymann RJ, Williams RE,
Magain P. 1992. {\it ApJ} 389:39

\noindent Smette A, Robertson JG, Shaver PA, Reimers D, Wisotzki L, Koehler T. 1995a.
{\it A\& A} 113:199

\noindent Smette A. 1995b. {\it QSO Absorption Lines}. p275. ed. Meylan G. Berlin: Springer
Verlag
\noindent Songaila A, Hu EM, Cowie LL. 1995. {\it Nature} 375:124

\noindent Songaila A, Cowie LL. 1996. {\it AJ} 112:335

\noindent Spitzer L. 1956. {\it ApJ} 124:20

\noindent Sramek RA, Weedman DW. 1978. {\it ApJ} 221:468

\noindent Steidel CC, Sargent WLW. 1987 {\it ApJ} 313:171

\noindent Steidel CC, Sargent WLW. 1987. {\it ApJ} 318:11

\noindent Steidel CC. 1995. in {\it QSO Absorption Lines}. p 139. ed. Meylan G. Berlin: Springer Verlag

\noindent Stocke JT, Shull JM, Penton S, Donahue M, Carilli C. 1995. {\it ApJ} 451:24

\noindent Stockton AN, Lynds CR. 1966. {\it ApJ} 144:451

\noindent Trevese D, Giallongo E, Camurani L. 1992. {\it ApJ} 398:491

\noindent Tytler D. 1987. {\it ApJ} 321:49

\noindent Tytler D. 1987. {\it ApJ} 321:69

\noindent Tytler D, Fan X-M, Burles S, Cottrell L, Davis C et al. 1995. in {\it QSO Absorption Lines}. p 289. ed. Meylan G. Berlin: Springer Verlag

\noindent Ulmer A. 1996. {\it ApJ} 473:110

\noindent Umemura M, Ikeuchi S. 1985. {\it ApJ} 299:583

\noindent van Gorkom J, Bahcall JN, Jannuzi B, Schneider DP. 1993. {\it AJ} 106:2213

\noindent van Gorkom J, Carilli CL, Stocke JT, Perlman ES, Shull JM. 1996. 
{\it AJ} 112:1397

\noindent Vishniac ET, Ostriker JP, Bertschinger E. 1985. {\it ApJ} 291:399

\noindent Vishniac ET, Bust GS. 1987. {\it ApJ} 319:14

\noindent Vogel SN, Weymann RJ, Rauch M, Hamilton T. 1995. {\it ApJ} 441:162  

\noindent Vogt SS, Allen SL, Bigelow BC, Bresee L, Brown B et al. 1994. {\it S.P.I.E.} 2198:362

\noindent Wadsley JW, Bond JR. 1997. in {\it Computational Astrophysics. Proc. 12th Kingston
Conference} Halifax Oct. 1996. p323. ed D. Clark, M. West. PASP conf. series.

\noindent Wagoner R. 1967. {\it ApJ} 149:465

\noindent Walsh D, Carswell RF, Weymann RJ. 1979. {\it Nature} 179:381

\noindent Wang B. 1995. {\it ApJ} 444:17

\noindent Webb JK. 1986. Proc. of the 124 IAU Symposium, p.803.
in (eds.):Hewitt, A., Burbidge G., Fang Z-F, Bejing 1986. Dordrecht:Reidel

\noindent Webb JK, 1987. PhD Thesis, Cambridge University.

\noindent Webb JK, Larsen I. 1988. in {\it High Redshift and Primeval Galaxies}.
eds. Bergeron J, Kunth D, Rocca-Volmerange B. Editions Fronti\`ere: Gif-sur-Yvette

\noindent Webb JK, Carswell RF. 1991. in Shaver PA, Wampler EJ, Wolfe AM, eds.
{\it Proceedings of the ESO Mini-Workshop on QSO Absorption Lines}
ESO Scientific Report No. 9,3. Garching: ESO

\noindent Webb JK, Barcons X. 1991. {\it MNRAS} 250:270

\noindent Webb JK, Barcons X, Carswell RF, Parnell HC. 1992. {\it MNRAS} 255:319

\noindent Weinberg DH, Miralda-Escud\'e J, Hernquist L, Katz N. 1997. {\it ApJ} 490:564

\noindent Weisheit JC 1978. {\it ApJ}, 219: 829

\noindent Weymann RJ, Carswell RF, Smith MG. 1981. {\it ARAA} 19:41

\noindent White SDM. 1979. {\it MNRAS} 186:145 

\noindent Williger GM, Carswell RF, Webb JK, Boksenberg A, Smith MG. 1989.
{\it MNRAS} 237:635

\noindent Williger GM, Babul A. 1992. {\it ApJ} 399:385 

\noindent Williger GM, Baldwin JA, Carswell RF, Cooke AJ, Hazard C, et al. 1994. {\it ApJ} 428:574

\noindent Wolfe AM, Turnshek DA, Smith HE, Cohen RD. 1986. {\it ApJ} 61:249

\noindent Womble DS, Sargent WLW, Lyons RS. 1996. in: {\it Cold Gas at High
Redshift} p. 137. (eds.) Bremer MN et al. Kluwer

\noindent Young PJ, Sargent WLW, Boksenberg A, Oke JB. 1981. {\it ApJ} 249:415

\noindent Young PJ, Sargent WLW, Boksenberg A. 1982. {\it ApJS} 48:455

\noindent Young PJ, Sargent WLW, Boksenberg A. 1982. {\it ApJ} 252:10

\noindent Zeldovich YB. 1970. {\it A\&A} 5:84

\noindent Zhang Y, Anninos P, Norman ML. 1995. {\it ApJLetters} 453:57 

\noindent Zhang Y, Anninos P, Norman ML, Meiksin A. 1997. {\it ApJ} 485:496

\noindent Zheng W, Davidsen AF, Kriss GA. 1998. {\it AJ} 115:391.

\noindent Zuo L. 1992. {\it MNRAS} 258:45

\noindent Zuo L. 1993. {\it A\& A} 278:343

\noindent Zuo L, Lu L. 1993. {\it MNRAS} 418:601

\noindent Zuo L, Bond JR. 1994. {\it ApJ} 423:73

\noindent 
\pagebreak

\pagebreak

\begin{figure}[t]
\centerline{
\psfig{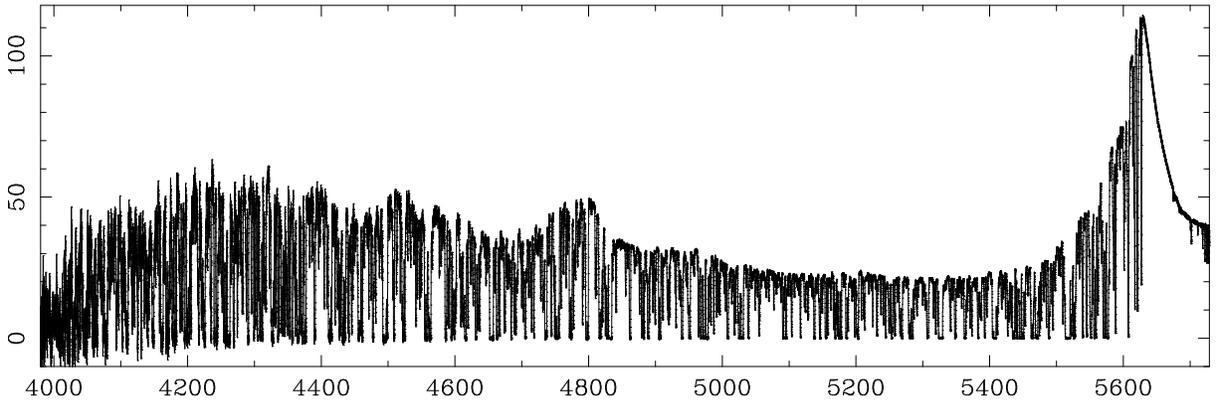}
}
\caption{High resolution (FWHM $\approx$ 6.6 kms$^{-1}$) spectrum 
of the $z_{em}$ = 3.62 QSO 1422+23 (V = 16.5), taken with the Keck
HIRES (signal-to-noise ratio $\sim$ 150 per resolution element, exposure time 25000 s).
Data from Womble et al  (1996).\label{spec}}
\end{figure}
\pagebreak

\begin{figure}[t]
\centerline{
\psfig{file=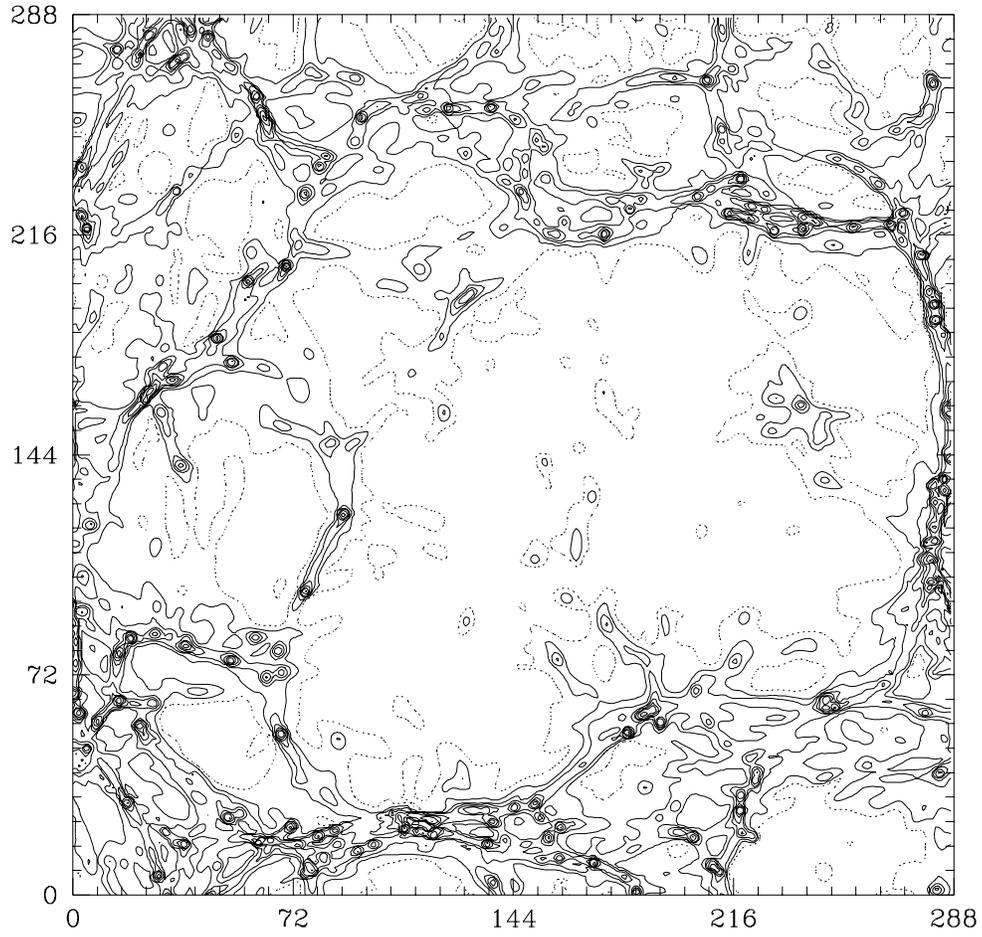,width=16.cm,angle=0.}
}
\caption{HI column density contours for a slice of the 10 h$^{-1}$ Mpc
(comoving) box from the $\Lambda$CDM simulation by Miralda-Escud\'e et al (1996)\label{bigpic}}
\end{figure}
\pagebreak

\begin{figure}[t]
\centerline{
\psfig{file=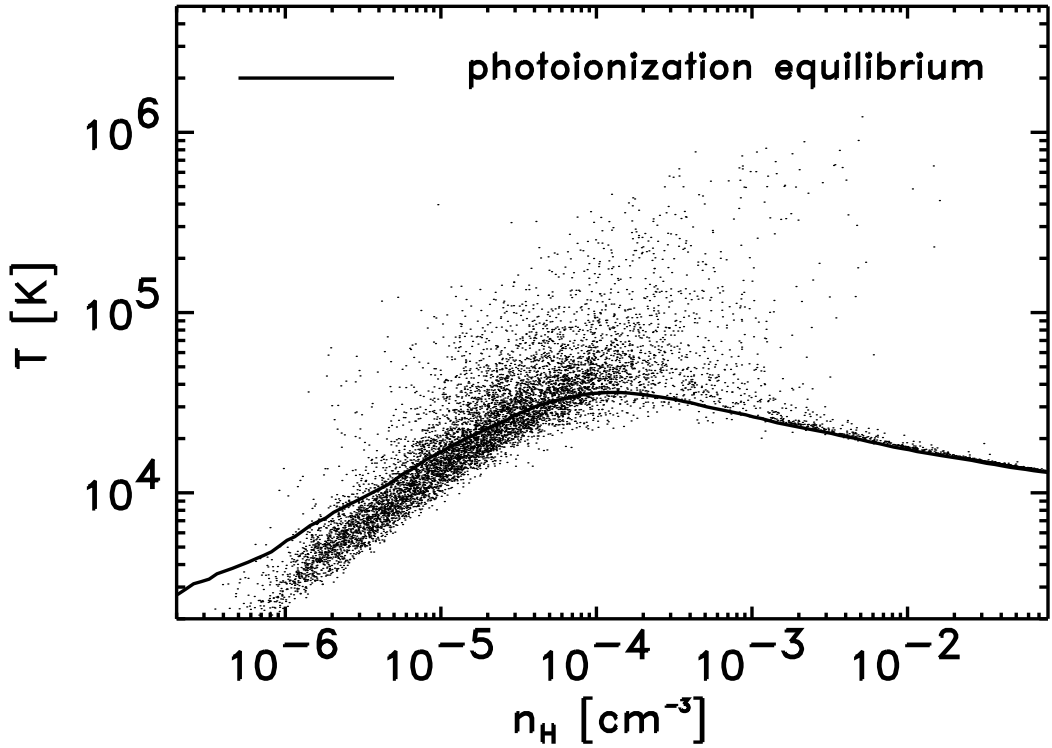,width=16.cm,angle=0.}
}
\caption{Density -- temperature ($n-T$) diagram of the
Ly$\alpha$ forest at redshift 3.1 from an SPH simulation of a standard
CDM universe (Haehnelt et al 1996b).  Each dot represents the mean
values of the total hydrogen density $n$ and the gas temperature $T$
along a random line of sight through the simulated box. The solid curve
gives the locus of thermal photoionization equilibrium. Departures from
this curve are due to the dynamical nature of the universe.  For all
but the most dense regions expansion cooling in voids (low density
regions) and heating by compression and shocks during gravitational
collapse steepen the $T(n)$ relation compared to the equilibrium curve
\label{nt}}
\end{figure}

\end{document}